\documentclass[fleqn,usenatbib]{mnras}

\usepackage{newtxtext,newtxmath}

\usepackage[T1]{fontenc}

\DeclareRobustCommand{\VAN}[3]{#2}
\let\VANthebibliography\thebibliography
\def\thebibliography{\DeclareRobustCommand{\VAN}[3]{##3}\VANthebibliography}

\usepackage{graphicx}	
\usepackage{amsmath}	
\usepackage{multirow}
\usepackage{xcolor}
\usepackage{ulem}

\newcommand{\Msun}{\,\mathrm{M}_\odot}
\newcommand{\feh}{\mathrm{[Fe/H]}}
\newcommand{\ttid}{t_{\mathrm{tid}}}

\newcommand{\Mgc}{M_{\mathrm{GC}}}
\newcommand{\Ngc}{N_{\mathrm{GC}}}
\newcommand{\Mtot}{M_{\mathrm{tot}}}
\newcommand{\Mh}{M_{\mathrm{h}}}
\newcommand{\Mstar}{M_{\star}}

\newcommand{\rms}{{\sc rms\ }}

\title[GC formation in dwarf galaxies]{Formation of globular clusters in dwarf galaxies of the Local Group}

\author[Y. Chen and O. Y. Gnedin]{Yingtian Chen\thanks{E-mail: ybchen@umich.edu} \href{https://orcid.org/0000-0002-5970-2563}{\includegraphics[scale=0.3]{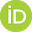}} and
Oleg Y. Gnedin \href{https://orcid.org/0000-0001-9852-9954}{\includegraphics[scale=0.3]{figures/orcid.png}}
\\
Department of Astronomy, University of Michigan, Ann Arbor, MI 48109, USA
}

\date{Accepted XXX. Received YYY; in original form ZZZ}

\pubyear{2023}

\begin{document}
\label{firstpage}
\pagerange{\pageref{firstpage}--\pageref{lastpage}}
\maketitle

\begin{abstract}
The existence of globular clusters (GCs) in a few satellite galaxies, and their absence in majority of dwarf galaxies, present a challenge for models attempting to understand the origins of GCs. In addition to GC presence appearing stochastic and difficult to describe with average trends, in the smallest satellite galaxies GCs contribute a substantial fraction of total stellar mass. We investigate the stochasticity and number of GCs in dwarf galaxies using an updated version of our model that links the formation of GCs to the growth of the host galaxy mass. We find that more than 50\% of dwarf galaxies with stellar mass $\Mstar\lesssim 2\times10^7\Msun$ do not host GCs, whereas dwarfs with $\Mstar\sim10^8\Msun$ almost always contain some GCs, with a median number $\sim 10$ at $z=0$. These predictions are in agreement with the observations of the Local Volume dwarfs. We also confirm the near-linear GC system mass--halo mass relation down to $\Mh\simeq10^8\Msun$ under the assumption that GC formation and evolution in galaxies of all mass can be described by the same physical model. A detailed case study of two model dwarfs that resemble the Fornax dwarf spheroidal galaxy shows that observational samples can be notably biased by incompleteness below detection limit and at large radii.
\end{abstract}



\section{Introduction}
\label{sec:intro}

Observations show a tight near-linear relation between the mass of a globular cluster (GC) system and the total mass of the host halo \citep{spitler_new_2009,georgiev_globular_2010,hudson_dark_2014,harris_dark_2015,forbes_extending_2018}. For example, \citet{harris_dark_2015} find $\Mgc=3.4\times10^{-5}M_{\rm h}$ for galaxies with halo mass between $10^{10}$ and $10^{14}\Msun$, with a total scatter of 0.35 dex most of which can be contributed by measurement uncertainties. Considering the complicated interplay of various non-linear baryonic processes involved in the formation of GCs, such a simple relation is quite remarkable. 

At a host mass $\Mh \sim 10^{9}\Msun$ the expected number of GCs is 1 or 0. In such a regime the cluster formation must become very stochastic. Therefore, it is particularly interesting to investigate how far the near-linear $\Mgc$--$M_{\rm h}$ relation holds. The main uncertainty is not the number of GCs but the measurement of the total halo mass of dwarf galaxies. \citet{forbes_extending_2018} used stellar and $\rm HI$ gas kinematics to derive dynamical mass measurements for dwarf galaxies in the Local Group (LG) and isolated late-type dwarfs with detected GC systems. They concluded that the number of GCs still correlates almost linearly with the halo mass down to $M_{\rm h}\lesssim 10^9\Msun$, although their derived halo masses fall systematically lower than those predicted by empirical stellar mass-halo mass (SMHM) relations such as those found by \citet{behroozi_average_2013} and \citet[][who investigated this relation in nearby dwarf galaxies]{danieli_elves_2022}.

Another challenge to study the $\Mgc$--$M_{\rm h}$ relation is small number of GCs in dwarf galaxies. Therefore, measuring the number of GCs can be heavily affected by incompleteness and contamination in surveys of dwarf galaxies. Fortunately, the Exploration of Local VolumE Satellites Survey \citep[ELVES,][]{carlsten_elves_2022,carlsten_exploration_2022} has extended the census of GC systems in a sample of 140 confirmed early-type dwarf satellite galaxies with stellar mass between $10^{5.5}$ and $10^{8.5}\Msun$. These authors parameterized and optimized the number of GCs as a function of stellar mass of the host galaxies $\Mstar$. For the low-mass regime where a significant fraction of galaxies do not host GCs, they calculated the occupation fraction (the fraction of galaxies hosting at least one GC) as a function of $\Mstar$. They found that the number of GCs increases monotonically with galaxy stellar mass, and the occupation fraction rises rapidly from 0 to 1 for galaxies with $\Mstar$ growing from $10^6$ to $10^8\Msun$. 

The ELVES survey does not provide direct measurement of host halo mass. Only a limited number of nearby dwarf galaxies have independent measurements of both halo mass and GC mass/number. This motivates the use of numerical methods to understand the formation of GCs, such as applying a model of GC formation and evolution to galaxy formation simulations such as the E-MOSAICS project \citep{pfeffer_e-mosaics_2018,kruijssen_e-mosaics_2019}, EMP-Pathfinder \citep{reina-campos_introducing_2022}, the model presented by \citet{doppel_modelling_2022}, and our previous models \citep{muratov_modeling_2010,li_modeling_2014,choksi_formation_2018,chen_modeling_2022}. These works have successfully reproduced the near-linear $\Mgc$--$M_{\rm h}$ relation in the mass range between $M_{\rm h}\sim10^{12}$ and $10^{14}\Msun$, without explicitly linking GC formation to the halo mass of host galaxies. However, \citet{choksi_formation_2018} noticed a departure from linearity at the low-mass end of $M_{\rm h}\sim10^{11}\Msun$. \citet{bastian_globular_2020} further extended the $\Mgc$--$M_{\rm h}$ relation relation down to $M_{\rm h}\sim10^{10}\Msun$ and found this relation to deviate downwards significantly below $M_{\rm h}\sim 5\times10^{11}\Msun$, in contrast to \citet{forbes_extending_2018} who found the near-linear correlation to be valid down to $M_{\rm h}\sim10^8\Msun$. The causes of the deviation in numerical works are still unclear. \citet{bastian_globular_2020} argued that this is because of the highly non-linear and uncertain SMHM relation at the low-mass end. Purely numerical limitations, such as inadequate mass resolution for dwarf galaxies, may also play a role.

Another caveat of numerical modelling is that most models cannot correctly reproduce the present-day cluster mass function from the assumed initial mass function, mainly because the treatment of tidal disruption is problematic. Due to the limited mass resolution in galaxy formation simulations, tidal disruption is normally modeled via subgrid prescriptions, which are unavoidably over-simplified. Moreover, the inadequate spatial resolution in simulations makes it challenging to explicitly calculate the tidal field on a scale of the tidal radius of GCs, $20-50$ pc. 

In this work, we apply our latest GC formation model presented in \citet{chen_modeling_2022} to a suite of higher resolution collisionless simulations, which are specifically tuned to the LG environment. The simulations have mass resolution of $2\times10^{5}\Msun$, enabling robust modelling of even the smallest dwarf galaxies down to $M_{\rm h}\sim10^8\Msun$. We modify the cluster sampling process in the model to make it work with dark matter (DM) particles. Also, we update the prescription for tidal disruption based on the most recent results of direct $N$-body simulations. We find our results consistent with the ELVES survey of the Local Volume (LV) GCs. We also investigate which aspects of the model can be constrained by the observational data. 

The paper is organized as follows. First, we recap the GC formation model in \citet{chen_modeling_2022} and introduce the modifications that we make to the model in Sec.~\ref{sec:model_setup}. Next, we present our main results in Secs.~\ref{sec:number_of_globular_clusters}, \ref{sec:detailed_example}, and \ref{sec:comparing_settings}. In Sec.~\ref{sec:number_of_globular_clusters}, we analyse the GC occupation fraction and the number/mass of GCs in the model galaxies with different stellar/halo mass. Next, we perform a detailed case study of the GC systems in two model galaxies that resemble the Fornax dwarf spheroidal galaxy (Fornax dSph) in Sec.~\ref{sec:detailed_example}. In Sec.~\ref{sec:comparing_settings}, we investigate how different model settings influence the GC occupation fraction and the number of GCs and constraint the models with observational data. We summarize our key findings in Sec.~\ref{sec:summary}.

\section{Model setup}
\label{sec:model_setup}

To investigate the formation of GCs in the LG dwarf galaxies, we apply our model of GC formation on a suite of cosmological simulations that resemble present-day properties of the LG environment. In this section, we describe the simulations and the GC model.

\subsection{Simulations of the Local Group}

We use a suite of collisionless (`DM only') zoom-in simulations with initial conditions (ICs) chosen to match the present-day LG. Full galaxy formation runs with these ICs are presented in \citet{brown_testing_2022}. The simulations are performed with the Adaptive Refinement Tree (ART) code \citep{kravtsov_adaptive_1997}. The ICs are \texttt{Thelma \& Louise} (in short, \texttt{T\&L}) and \texttt{Romeo \& Juliet} (\texttt{R\&J}). The modifications from the original version of \citet{garrison-kimmel_elvis_2014} include reducing the simulation box sizes to $\sim 35$ comoving Mpc and improving the root grid resolution \citep{brown_improving_2021}. The zoom-in region is around 10 comoving Mpc across, and the particle mass in the zoom-in region is smaller than $2\times10^5\Msun$. We summarize the key parameters for the two ICs in Table~\ref{tab:IC}. 

\begin{table*}
 \caption{Important simulation parameters of the initial conditions employed in this work. The lengths are given in comoving units and the particles mass refers to the particles in the zoom-in region. The $z=0$ halo mass of the two main galaxies in each IC are also given.}
 \label{tab:IC}
 \renewcommand\arraystretch{1.2}
 \begin{tabular}{cccccccccc}
  \hline
  IC & Box size & Root cell size & \# of particles & Particle mass & $\Omega_{\rm m}$ & $M_{\rm h,1}$ & $M_{\rm h,2}$ & $\Omega_{\rm b}$ & $h$ \\
  \hline
  \texttt{T\&L} & 35.2 Mpc & 138 kpc & 65,589,112 & $1.89\times10^5\Msun$ & 0.266 & $1.09\times10^{12}\Msun$ (\texttt{T}) & $0.94\times10^{12}\Msun$ (\texttt{L}) & 0.0449 & 0.71 \\
  \texttt{R\&J} & 34.0 Mpc & 133 kpc & 56,765,377 & $1.82\times10^5\Msun$ & 0.31 & $1.28\times10^{12}\Msun$ (\texttt{R}) & $0.97\times10^{12}\Msun$ (\texttt{J}) & 0.048 & 0.68 \\
  \hline
 \end{tabular}
\end{table*}

We start the simulation at $z\simeq100$ and run it until the present. We output simulation snapshots at approximately every $0.01$ increment of the scale factor $a$. Next, we generate halo catalogues at each snapshot with the \textsc{rockstar} halo finder \citep{behroozi_rockstar_2013}. The halo catalogues and simulation snapshots are then passed to the \textsc{consistent tree} code \citep{behroozi_gravitationally_2013} to generate merger trees. 

The mass assembly of the four main galaxies in the two ICs can be split into two categories \citep[see Fig.~3 in][for the mass growth histories]{brown_testing_2022}. \texttt{Louise}, \texttt{Romeo}, and \texttt{Juliet} have no major merger with a mass ratio less than 4:1 after $z\simeq 5$, which resembles the formation history of the Milky Way (MW) \citep{hammer_milky_2007}. We therefore refer to the three galaxies as `MW-like'. In contrast, the \texttt{Thelma} galaxy encounters more major mergers at later times.

\subsection{Modelling the formation and evolution of globular clusters}
\label{sec:GC_formation_model}

We apply a GC formation model on the simulation outputs to study GC systems of the LG galaxies. Based on \citet{chen_modeling_2022}, we describe GC formation and evolution via four steps: 1) cluster formation, 2) cluster sampling, 3) particle assignment, and 4) cluster evolution. In this section, we recap the GC model and describe several modifications required to study dwarf galaxies.

\subsubsection{Cluster formation}

In the cluster formation step, we trigger a GC formation event when the specific mass accretion rate of the host galaxy exceeds a threshold value, $p_3$, which is an adjustable model parameter. The specific mass accretion rate, $R_{\rm m}$, is defined as the fractional change of galaxy mass between two adjacent simulation snapshots:
\begin{equation}
	R_{\rm m} = \frac{M_{\rm now} - M_{\rm prog}}{M_{\rm prog}}\cdot\frac{1}{t_{\rm now}-t_{\rm prog}}
	\label{eq:mass_accretion_rate}
\end{equation}
where $t_{\rm now}$ and $t_{\rm prog}$ stand for the cosmic times of the current snapshot and the progenitor snapshot, respectively. Similarly, the masses of the current galaxy and the main progenitor galaxy are represented by $M_{\rm now}$ and $M_{\rm prog}$. Since the mass of DM particles in zoom-in regions is around $2\times 10^5\Msun$ we only take into account halos with $M_{\rm h}>10^8\Msun$ to ensure that each halo contains at least 500 particles. Halos smaller than that may be numerically under-resolved, but they are very unlikely to host any massive star clusters.

When a cluster formation event is triggered, we analytically calculate the stellar mass of a galaxy from its halo mass using the SMHM relation proposed by \citet{behroozi_average_2013}, with a redshift-dependent scatter $\xi(z)=0.218+0.0203z/(1+z)$. We then follow \citet{choksi_formation_2018} to evolve the stellar mass self-consistently. First, we assign an initial stellar mass to the first progenitor along each branch, $\Mstar^0$, sampled from a Gaussian distribution, ${\cal N}(\overline{\Mstar^0},\xi(z_0))$. The average value $\overline{\Mstar^0}$ refers to the raw stellar mass from SMHM without scatter. Next, we evolve the stellar mass as
\begin{equation*}
    \Mstar^{\rm now} = \Mstar^{\rm prev} + \left(\overline{\Mstar^{\rm now}}-\overline{\Mstar^{\rm prev}}\right)10^{{\cal N}(0,\xi(z_{\rm now}))}.
\end{equation*}
This method preserves some memory of the historical stellar mass, so that a galaxy deviating from the mean SMHM at the beginning tends to continue the trend.

Using the stellar mass, we calculate the cold gas mass via the gas mass--stellar mass relation by \citet{choksi_formation_2018}:
\begin{equation}
    \eta(\Mstar, z)=\frac{M_{\rm g}}{\Mstar}=0.35\times 3^{2.7}\left(\frac{\Mstar}{10^9\Msun}\right)^{-n_M(\Mstar)}\left(\frac{1+z}{3}\right)^{n_z(z)}
    \label{eq:gas_mass_stellar_mass}
\end{equation}
based on the observations of \citet{lilly_gas_2013,genzel_combined_2015,tacconi_phibss_2018,wang_3_2022}. Here $n_M(\Mstar) = 0.33$ for $\Mstar > 10^9 \Msun$ and $n_M(\Mstar) = 0.19$ for $\Mstar < 10^9 \Msun$. The redshift dependency is characterized by $n_z(z) = 1.4$ for $z>2$ and $n_z(z) = 2.7$ otherwise. When $z>3$, following \citet{li_modeling_2014} we adopt a fixed upper limit: $\eta(\Mstar, z>3) = \eta(\Mstar, z=3)$. An intrinsic scatter of $0.3$ dex is also added to this relation. 

Another constraint on the gas mass of the host galaxy is that sum of the gas fraction $f_{\rm g}=M_{\rm g}/M_{\rm h}$ and the stellar fraction \mbox{$f_*=\Mstar/M_{\rm h}$} cannot exceed the total accreted baryon fraction $f_{\rm in}$, which is limited by extragalactic UV background after reionization. Since this condition is particularly important for dwarf galaxies, here we update the expression for $f_{\rm in}$ used in our previous models since \citet{muratov_modeling_2010}. The new expression from \citet{kravtsov_span_2022} takes the form
\begin{equation}
    f_{\rm in} = f_{\rm b} \, s(M_{\rm ch}(z)/M_{\rm h}, 2),
    \label{eq:baryon_fraction}
\end{equation}
where $f_{\rm b}=\Omega_{\rm b}/\Omega_{\rm m}$ is the universal baryon fraction, $s(x,y) = [1+(2^{y/3}-1)x^y]^{-3/y}$ is a soft step-function, and $M_{\rm ch}$ is the characteristic mass scale at which $f_{\rm in}=0.5f_{\rm b}$:
\begin{equation}
    M_{\rm ch}(z) = 1.69\times 10^{10}\Msun\frac{\exp({-0.63z})}{1+\exp[(z/\beta)^\gamma]},
    \label{eq:mass_scale}
\end{equation}
where
\begin{equation}
    \beta = z_{\rm rei}\left[\ln\left(1.82\times 10^3\exp(-0.63z_{\rm rei})-1\right)\right]^{-1/\gamma}.
    \label{eq:beta}
\end{equation}
We adopt the reionization epoch at $z_{\rm rei} = 6$ and $\gamma=15$ as in \citet{kravtsov_span_2022}. If $f_{\rm g}+f_* > f_{\rm in}$, we set $f_{\rm g} = f_{\rm in}-f_*$. The new expression of $f_{\rm in}$ is similar to the one in our previous models at $z\lesssim 4$, but gives significantly larger values at higher redshift. Such a constraint is important for halos with $M_{\rm h} \lesssim 10^9\Msun$ at $z\simeq2$ when the formation of GCs is active.

The linear cluster mass--gas mass relation obtained from a simulation by \citet{kravtsov_formation_2005} is employed to calculate the total mass of a newly formed GC population:
\begin{equation}
	\Mtot = 1.8\times 10^{-4} \, p_2 \, M_{\rm g}
	\label{eq:total_cluster_mass}
\end{equation}
where $M_{\rm g}$ is the cold gas mass of the host galaxy, and $p_2$ is another adjustable parameter\footnote{For consistency with previous work, we keep the notation of $p_2$ and $p_3$ as in \citet{li_modeling_2014}.}. This linear relation intuitively links the intensity of cluster formation to the total gas mass of the host galaxy, reflecting the fact that star clusters are formed in gas clouds \citep{shu_star_1987,scoville_far-infrared_1989,mckee_theory_2007,krumholz_star_2019}. Similar relation is also observed in elliptical galaxies by \citet{mclaughlin_efficiency_1999}, who found that the ratio between cluster mass and baryon mass is roughly a constant.

The metallicity of the newly formed cluster population is directly drawn from the metallicity of the interstellar medium of the host galaxy, which is given by
\begin{equation}
	\feh=\log \left[\left(\frac{\Mstar}{10^{10.5}\Msun}\right)^{0.35}(1+z)^{-0.9}\right].
	\label{eq:metalicity_stellar_mass}
\end{equation}
We follow \citet{ma_origin_2016} to employ $0.35$ slope for the stellar mass dependency. The $0.9$ slope of the redshift dependency is calculated based on the observations of Lyman-break galaxies by \citet{mannucci_lsd_2009}, who found a 0.6 dex drop of $\feh$ from $z=0$ to $\sim4$. In addition, we apply a $0.3$ dex intrinsic scatter to $\feh$.

\subsubsection{Cluster sampling}
\label{sec:cluster_sampling}

The next step is GC sampling, where we compute an initial mass of each individual cluster. For each GC formation event with total mass $\Mtot$, we sample the masses of individual clusters from a \citet{schechter_analytic_1976} initial cluster mass function (ICMF) with a power-law slope of $-2$:
\begin{equation}
    \frac{dN}{dM} \propto M^{-2}e^{-M/M_{\rm c}}.
    \label{eq:schechter}
\end{equation}
Following \citet{choksi_formation_2019}, we set $M_{\rm c}=10^7\Msun$. To numerically draw clusters from the ICMF, we first calculate the cumulative distribution function
\begin{equation}
    r(M)\equiv\frac{N(<M)}{N(<M_{\rm max})} = \frac{\int_{M_{\rm min}}^{M}\frac{dN}{dM}dM}{\int_{M_{\rm min}}^{M_{\rm max}}\frac{dN}{dM}dM},
    \label{eq:cumulative}
\end{equation}
where $M_{\rm min/max}$ are the minimum/maximum cluster mass. We will specify the selection of $M_{\rm min/max}$ later. The variable $r(M) \in [0,1)$ is a monotonic function for any $M\in[M_{\rm min},M_{\rm max})$, and thus $r(M)$ is invertible. Then, we draw a random number $x\in [0,1)$ and convert $x$ to a cluster mass via $M=r^{-1}(x)$. We repeat the process until the total mass of newly formed clusters, $\Mgc$, just exceeds $\Mtot$. We drop the last cluster (with mass $M$) with a probability $P=(\Mgc-\Mtot)/M$. Therefore, the expected value of $\Mgc$ is $E(\Mgc)=\Mgc(1-P) + (\Mgc-M)P=\Mtot$.

In a rare case of $\Mtot<M_{\rm min}$, we still randomly draw a cluster from the ICMF with the above method. However, since the mass of such a cluster, $M$, is greater than $M_{\rm min}$ (and thus greater than $\Mtot$), we must stochastically determine whether to keep it to ensure that the expected value of $\Mgc$ is still $\Mtot$. Therefore, we keep this cluster with a probability $P=\Mtot/M$, so that the expected value of $\Mgc$ is $E(\Mgc)=MP=\Mtot$. By employing these techniques, we can guarantee that the expected value of $\Mgc$ is always $\Mtot$.

While in previous versions of our model we used the minimum cluster mass of $10^5\Msun$, here we set $M_{\rm min} = 10^4\Msun$ so that we can correctly model the masses of GCs even in the smallest halos with $M_{\rm h}\gtrsim 10^8\Msun$. We show our motivation with an order-of-magnitude calculation: plugging $M_{\rm g}\sim f_{\rm b}M_{\rm h} \gtrsim 10^7\Msun$ and $p_2\sim 10$ into equation~(\ref{eq:total_cluster_mass}), we get $\Mtot\gtrsim 10^4\Msun$. Therefore, we expect $M_{\rm min}\geq10^4\Msun$ to avoid the abnormal case of $\Mtot<M_{\rm min}$. However, clusters with $M<10^4\Msun$ will be disrupted relatively quickly by the tidal field: the estimated lifetime of $10^4\Msun$ cluster at 3 kpc from the galactic center of a MW-mass galaxy is less than 1 Gyr. Moreover, since we will mainly compare our results with the ELVES survey \citep{carlsten_elves_2022}, which is magnitude-limited to $M_{\rm g}<-5.5$, corresponding to $M>3\times10^4\Msun$, it is unnecessary to model clusters less massive than $10^4\Msun$. We thus set $M_{\rm min}=10^4\Msun$. Note that we adopted $M_{\rm min}=10^5\Msun$ in \citet{chen_modeling_2022} due to the limited mass resolution in that work. Therefore, the $p_2$ values in the two works are not directly comparable as a larger $p_2$ is needed to maintain the same ICMF at the high-mass end if we reduce $M_{\rm min}$ from $10^5\Msun$ to $10^4\Msun$.

In \citet{choksi_formation_2019} and \citet{chen_modeling_2022}, $M_{\rm max}$ is set to match the deterministic constraint
\begin{equation}
    1 = \int_{M_{\rm max}}^{\infty} \frac{dN}{dM}dM.
    \label{eq:mmax_constraint}
\end{equation}
In other words, there is only one cluster with $M=M_{\rm max}$ in each GC formation event. Such cluster is drawn first in the list of newly formed GCs. An alternative method is to set $M_{\rm max}\rightarrow\infty$ and allow more massive clusters to form with a small but non-zero probability. Numerically, we can set $M_{\rm max}$ to be a large enough number $\gtrsim 10^7\Msun$. These two methods produce similar results for galaxies with $M_{\rm h}\gg 10^9\Msun$. However, for galaxies with $M_{\rm h}\sim 10^9\Msun$, the former method prevents the formation of massive clusters ($M\sim 10^5\Msun$) as $M_{\rm max}$ is too small, whereas the latter method can still form massive clusters with nonzero probability. This may lead to noticeable difference in the final GC abundances in dwarf galaxies. In the rest of the work, we investigate both methods but treat $M_{\rm max}\rightarrow\infty$ as the default.

\subsubsection{Particle assignment}

After determining the masses of individual GCs, we link the newly formed clusters to collisionless particles in the simulation snapshots. This step is different from our previous work \citep{chen_modeling_2022}, where we applied the model mostly on stellar particles in the Illustris TNG50 simulation. There we chose particles representing young stellar populations, with age typically less than 10 Myr, that indicate likely formation sites of giant molecular clouds within the galaxy. In this work we have only DM particles from our new LG simulations, and therefore we need to assign GCs to DM particles near possible locations of giant molecular clouds. To try to find reasonable proxies for the cloud location, we search for peaks of matter density. These peaks may correspond to surviving dense cores of satellite galaxies or other galactic structure with deep potential wells. Giant gas clouds are more likely to be formed around such peaks than elsewhere, and we therefore adopt these local density peaks to mimic the location of giant clouds. We then identify local density peaks within $r_{\rm s}$ of the galaxy center, where $r_{\rm s}$ is the scale radius of the best-fit Navarro–Frenk–White (NFW) halo profile.We require the peak density to be higher than the mass density of the 16 closest grid cells and $30$ times the mean density enclosed within the $r_{\rm s}$ sphere. The first criterion ensures that the peak is located at a local maximum, while the second criterion excludes low-density peaks that are unlikely to host massive star clusters. The factor $30$ is chosen such that the resulting radial number density profiles of model GCs can match the observed profiles of both the MW and satellites.

To find all such density peaks, we start with the central peak and search for the next highest density peak outside $1$ kpc of the first peak. We repeat the process to search for the remaining peaks outside $1$ kpc of all existing peaks. Every time we find a peak, we assign one GC to the DM particle located near the center of the peak until we find all peaks satisfying the criteria or we have assigned all GCs to peaks. If there are more GCs than the number of peaks, we loop through the peaks again: first assign one GC to a random DM particle within $500$ pc of the first peak, then another GC to the second peak, and so on. We repeat the process until we have assigned all GCs to the peaks. This guarantees that each peak has approximately equal number of GCs. Benefited by the high mass resolution of the simulations, we can always find sufficient number of DM particles satisfying the above criteria even during the early epochs of galaxy formation.

This particle assignment ensures that the GC profiles of the three MW-like galaxies are consistent with the observed GC profile of the MW. After calibration (described in Sec.~\ref{sec:selecting_model_parameters}), our model gives the GC half number radius for \texttt{Louise}, \texttt{Romeo}, and \texttt{Juliet} to be $5.8$, $4.1$, and $4.3$~kpc, respectively, in agreement with the observed half number radius around $4.8$~kpc. We also notice that the projected GC profiles of the three MW-like galaxies have a flat core within the central $1$ kpc and follow a power-law function at $R=1-100$ kpc, with slopes varying from $-2.3$ to $-2.5$, being consistent with the $-2.4$ slope of the MW. In addition, the new assignment method allows GCs to form farther away from the galactic center. This assignment typically selects DM particles $200-5000$ kpc from the galactic center for $\Mh\sim10^{10}\Msun$, in agreement with \citet{sameie_formation_2023}, who employed hydrodynamic simulations and suggested that clusters are formed $\sim1000$ pc to the galactic center for $\Mh\sim10^{10}\Msun$.

\subsubsection{Cluster evolution}
\label{sec:cluster_evolution}

The final step is to follow the trajectories of GC particles and model the evolution of GC mass until the present. We take into account two main processes of mass evolution: tidal disruption and stellar evolution. The tidal disruption rate of a cluster with mass $M$ can be expressed as
\begin{equation}
	\frac{dM(t)}{dt} = -\frac{M(t)}{\ttid(M,t)}
	\label{eq:mass_loss_rate}
\end{equation}
where $\ttid$ is the tidal disruption timescale. As suggested by \citet{gieles_lifetimes_2008}, the disruption time depends significantly on the local tidal field strength, parametrized by the effective angular frequency $\Omega_{\rm tid}$. In the previous versions of this model \citep{choksi_formation_2018,chen_modeling_2022}
we used the expression
\begin{equation}
	t_{\rm tid}(M,t) = 10\, {\rm Gyr} \left[\frac{M(t)}{2\times10^5\Msun}\right]^{y}  \left[\frac{\tilde{\Omega}_{\rm tid}(t)}{100\ {\rm Gyr^{-1}}}\right]^{-1}
	\label{eq:t_tid}
\end{equation}
with $y=2/3$ and the tidal frequency estimated as $\tilde{\Omega}_{\rm tid}^2 \simeq \max(|\lambda_1|, |\lambda_2|, |\lambda_3|)/3$, where $(\lambda_1, \lambda_2, \lambda_3)$ are the three eigenvalues of tidal tensor ${\bf T}(\mathbfit{x}_0,t)$ sorted in descending order. This mass loss rate can be rewritten as
\begin{equation}
	\frac{dM(t)}{dt} = -20\, \frac{\Msun}{\rm Myr} \left[\frac{M(t)}{2\times10^5\Msun}\right]^{1-y}  \left[\frac{\tilde{\Omega}_{\rm tid}(t)}{100\ {\rm Gyr^{-1}}}\right].
\end{equation}

In this work, we apply a modified expression for the cluster mass loss, motivated by a re-evaluation of direct $N$-body models of cluster disruption by \citet{gieles_mass-loss_2023}:
\begin{equation}
  \frac{dM(t)}{dt} = -20\, \frac{\Msun}{\rm Myr} \left[\frac{M_{\rm i}}{2\times10^5\Msun}\right]^{1-x} \left[\frac{M(t)}{M_{\rm i}}\right]^{1-y} \left[\frac{\Omega_{\rm tid}(t)}{150\ {\rm Gyr^{-1}}}\right]
  \label{eq:disrupt}
\end{equation}
with potentially different scalings $x$ and $y$. The main change here is separating the overall normalization of the mass loss rate as a function of initial cluster mass $M_{\rm i}$ (via $x$) and the dependence on current cluster mass $M(t)$ (via $y$). We obtain the previous prescription if $x=y$ and $M_{\rm i}$ cancels out. However, recent $N$-body models indicate that the evolution slope $y$ may deviate from the initial $x$ depending on cluster density and even exceed the value of 1.

To explore systematic variation of our results on the modeling of tidal disruption, we consider three alternative models. The first is the old version, $x=y=2/3$. The second is a modified version with $x=y=1$, which should produce stronger disruption of low-mass clusters. The third model has $x=2/3$ but $y=4/3$, which is preferred by the new $N$-body models. This parametrization should also reduce the fraction of low-mass clusters.

\begin{figure}
    \includegraphics[width=\linewidth]{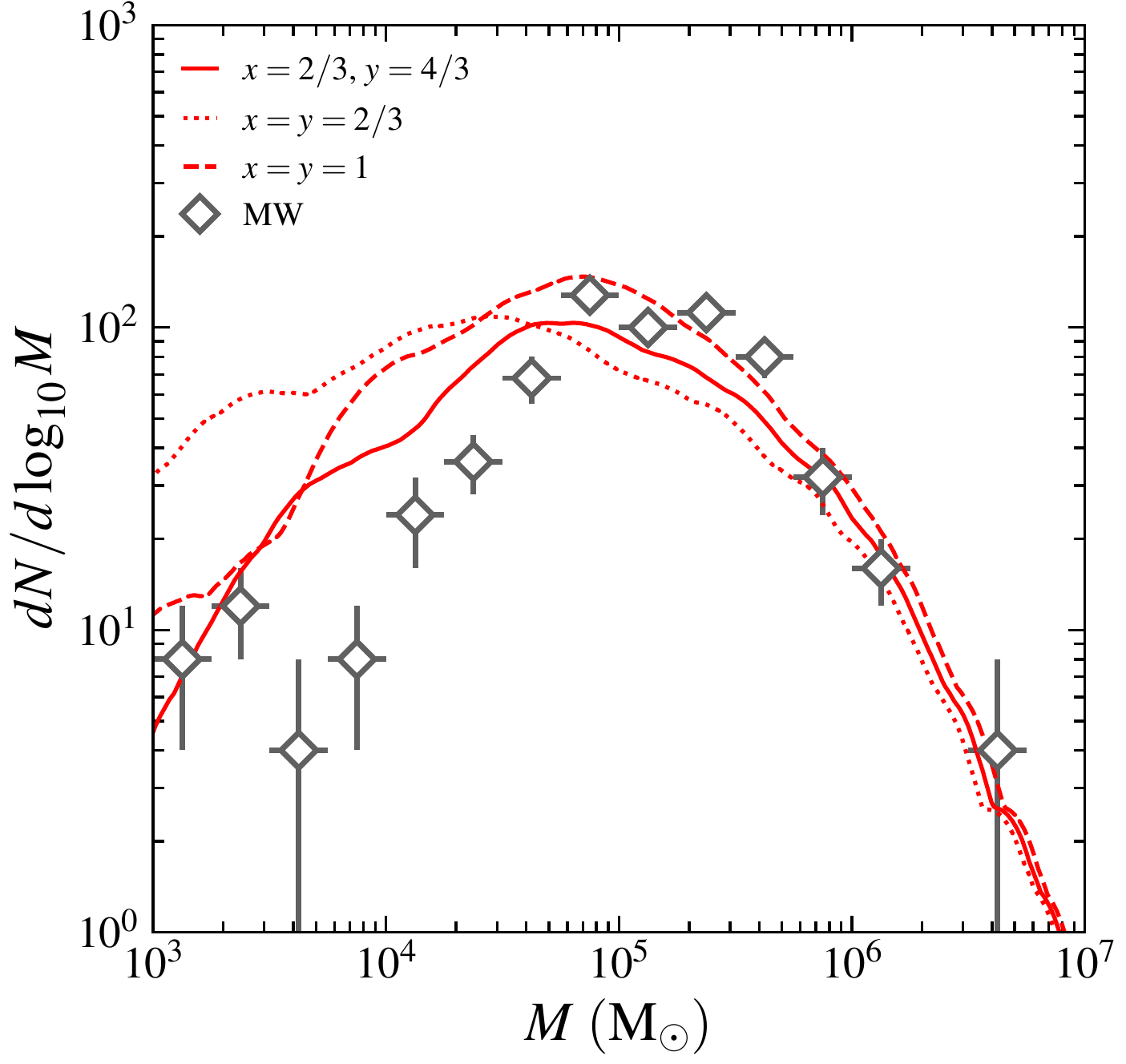}
    \vspace{-4mm}
    \caption{Average GC mass functions of the three MW-like galaxies with different prescriptions for tidal disruption, given by equation~(\ref{eq:disrupt}). Solid line is for $x=2/3,y=4/3$, dotted line is for $x=y=2/3$, and dashed line is for $x=y=1$. For comparison, the mass function of the MW GC system is overplotted as diamonds with errorbars: vertical errorbars show the interquartile ranges computed via bootstrap resampling, and horizontal errorbars correspond to the bin width. We repeat bootstrap resampling until the estimated interquartile ranges converge.}
   \label{fig:mass_function}
\end{figure}

The present-day GC mass function depends noticeably on the disruption models. In Fig.~\ref{fig:mass_function}, we compare the average mass functions of GCs in the three MW-like galaxies produced by the three models of tidal disruption. The model parameters are calibrated as we describe in Sec.~\ref{sec:selecting_model_parameters}. The mass function of the $x=2/3,y=4/3$ model lies between the two other models for $M>3\times10^4\Msun$ and predicts lower abundance of clusters below this mass, better matching the observed mass function of the MW GCs. Therefore, we treat the $x=2/3,y=4/3$ prescription as the default, and the other versions as alternates unless mentioned specifically.

We also use an updated expression for the tidal frequency $\Omega_{\rm tid}$ via the effective eigenvalue $\lambda_{\rm 1,e}$ that takes into account the centrifugal, Euler, and Coriolis forces \citep{renaud_evolution_2011}:
\begin{equation}
	\Omega_{\rm tid}^2\simeq\lambda_{\rm 1,e}\simeq\lambda_1-\lambda_3.
	\label{eq:omega_tid}
\end{equation}
Typically, $\lambda_1>0$ and $\lambda_3<0$. This expression reflects the mass loss more accurately than $\tilde{\Omega}_{\rm tid}$. The resulting values of $\Omega_{\rm tid}$ are systematically higher by a factor $1.2-2.1$ at $z=2-5$ when GC formation is the most active, so that we updated the normalization factor from 100 to 150~Gyr$^{-1}$ to maintain consistency with the previous versions of the model.

The tidal tensor is defined as
\begin{equation}
	T_{ij}(\mathbfit{x}_0,t) \equiv -\left.\frac{\partial^2\Phi(\mathbfit{x},t)}{\partial x_i\partial x_j}\right|_{\mathbfit{x}=\mathbfit{x}_0}
	\label{eq:tidal_tensor}
\end{equation}
where $i$ and $j$ are the orthogonal directions in the Cartesian coordinate system, and $\mathbfit{x}_0$ stands for the location of the cluster. Based on numerical experiments, \citet{pfeffer_e-mosaics_2018} suggested to use $\lambda_{\rm 1,e}\simeq\lambda_1-0.5(\lambda_2+\lambda_3)$. In spherical symmetry we have $\lambda_2=\lambda_3$, which leads to the second equality in equation~(\ref{eq:omega_tid}). 

We numerically calculate the tidal tensor by placing a $3\times 3\times 3$ cell cube centered on the GC particle, where the side length of the cell is $d$. We approximate the diagonal terms of the tidal tensor via 
\begin{equation}
    T_{ii}=-\frac{1}{d^2}[\Phi(\mathbfit{x}_0+\hat{\mathbfit{e}}_id)
    +\Phi(\mathbfit{x}_0-\hat{\mathbfit{e}}_id) 
    -2\Phi(\mathbfit{x}_0)],
    \label{eq:tii}
\end{equation}
where $\hat{\mathbfit{e}}_i$ is the unit vector along the $i$ direction. Similarly, the non-diagonal terms are given by
\begin{multline}
    T_{ij}=-\frac{1}{4d^2}[\Phi(\mathbfit{x}_0+\hat{\mathbfit{e}}_id+\hat{\mathbfit{e}}_jd)
    +\Phi(\mathbfit{x}_0-\hat{\mathbfit{e}}_id-\hat{\mathbfit{e}}_jd) \\
    -\Phi(\mathbfit{x}_0+\hat{\mathbfit{e}}_id-\hat{\mathbfit{e}}_jd)
    -\Phi(\mathbfit{x}_0-\hat{\mathbfit{e}}_id+\hat{\mathbfit{e}}_jd)].
    \label{eq:tij}
\end{multline}
As in \citet{chen_modeling_2022}, we set $d=300\ {\rm pc}$ for best accuracy in the regions containing most GCs. Although this value is still too large compared to the tidal radius of GCs ($20-50\,$pc) we cannot adopt a lower $d$ as we are limited by the spatial resolution of the simulation. In addition, since we apply the model on collisionless simulations, we cannot directly model the gravitational potential of baryons, which may be different from that of DM. Ignoring baryonic structure typically tends to underestimate the tidal force. However, this effect is not obvious for dwarf galaxies $\Mstar < 10^8\Msun$, which are dominated by DM. To correct the underestimate of the tidal field strength $\Omega_{\rm tid}$ due to both aforementioned effects, we boost it by the third adjustable model parameter $\kappa$:
\begin{equation}
	\Omega_{\rm tid}^2 = \kappa (\hat{\lambda}_1-\hat{\lambda}_3).
\end{equation}
The notation $\hat{\lambda}_i$ stands for the $i$-th eigenvalue of the tidal tensor calculated by the finite differences in equations~(\ref{eq:tii}) and (\ref{eq:tij}).

Using equation~(\ref{eq:disrupt}) we calculate the current mass of a GC at time $t$ after formation due to tidal disruption as $M'(t)$. Assuming the timescale of stellar evolution is much shorter than $t_{\rm tid}$, the final mass of the GC is given by
\begin{equation}
	M(t) = M'(t)\left[1-\int_0^t\nu_{\rm se}(t')\ dt'\right],
	\label{eq:stellar_evolution}
\end{equation}
where $\nu_{\rm se}$ is the mass loss rate due to stellar evolution by \citet{prieto_dynamical_2008}.

\subsection{Selecting model parameters}
\label{sec:selecting_model_parameters}

The model has three adjustable parameters $(p_2,p_3,\kappa)$ controlling the formation rate, formation timing, and disruption rate of GCs. To obtain the values of these parameters that match best the three MW-like galaxies (\texttt{Louise}, \texttt{Romeo}, and \texttt{Juliet}), we compare several key properties of surviving clusters with the observational data of MW GC system, including the number of clusters, mass function, metallicity distribution, and radial profile. We calibrate the model specifically for the MW because the observations of the MW GC system are the most complete among all GC systems. This allows comparison of the model GC systems with observations in many different aspects, as we introduce below. Also, the mass assembly history of the MW is understood better than any other galaxy of similar masses, such as M31. Since the mass assembly history is one of the key inputs of the model, we are more confident that the three MW-like galaxies should produce GC systems similar to the MW GC system with the calibrated model parameters.
The calibration is done by minimizing the following merit function:
\begin{equation}
    {\cal M} = \chi^2_N + \chi^2_\sigma + G_M + G_Z + G_R.
    \label{eq:merit}
\end{equation}

The first term is the reduced $\chi^2$ of the number of surviving clusters:
\begin{equation}
    \chi^2_N = \frac{1}{N_{\rm h}}\sum_{i=1}^{N_{\rm h}}\frac{(\log N_{i}-\log N_{\rm MW})^2}{\sigma_{\log N}^2},
\end{equation}
where $N_{\rm h}=3$ is the number of MW-like galaxies in our simulations, $N_{i}$ is the number of surviving clusters in the $i$-th simulated galaxy, and $N_{\rm MW}=150$ in the observed number of GCs in the MW. We adopt Poisson's error of $\sigma_{\log N}=0.04$.

Similarly, the second term is the reduced $\chi^2$ of the velocity dispersion of surviving clusters. Here the velocity dispersion is defined as the 3D dispersion, $\sigma^2 \equiv \sigma_R^2 + \sigma_\phi^2 + \sigma_z^2$, for the MW and its simulated analogues.

The remaining three terms in equation~(\ref{eq:merit}) are the `goodness' of the mass function, metallicity distribution, and radial profile, respectively. For example, $G_M$ stands for the inverse of the fraction of MW-like galaxies that can match the observed mass function of surviving GCs. By performing the Kolmogorov–Smirnov (KS) test on the model galaxies with observations, we define a galaxy to match observation if the $p$-value of KS test exceeds 0.01. Similarly, $G_Z$ is the inverse of the fraction of MW-like galaxies that can match the observed distribution of $\feh$, and $G_R$ is the inverse of the fraction of MW-like galaxies that can match the observed radial profile, i.e., the distribution of face-on projected distance between GCs and the galaxy center.

By minimizing the merit function, we find the best parameters to be $(p_2,p_3,\kappa) = (14,0.7\ {\rm Gyr^{-1}}, 1.5)$ both for the default prescription of tidal disruption, $x=2/3,y=4/3$, and for the case of $x=y=1$. For the old model with $x=y=2/3$, the best parameters are $(p_2,p_3,\kappa) = (14,0.7\ {\rm Gyr^{-1}}, 2.5)$. It is worth emphasizing that these parameter sets are calibrated specifically for the MW system. The results for satellite galaxies are true predictions of the model.

\subsection{Selecting dwarf galaxies}

The main goal of this work is to investigate the formation of GCs in dwarf galaxies, especially the satellite galaxies that are associated with MW and M31. To achieve this, we select 10 satellite galaxies with the highest maximum halo mass for either \texttt{T\&L} and \texttt{R\&J}, yielding 20 satellite galaxies in total. We define satellite galaxy as the galaxy located inside the virial radius of any of the main galaxies at $z=0$. The `highest maximum halo mass' refers to the mass of the historically most massive progenitor galaxy in the merger tree. We apply the model on this halo sample and analyse the GC systems in these galaxies throughout the paper.

Since most dwarf galaxies have only a few or even no GCs, the model randomness can play an important role in shaping the GC systems. The randomness includes the scatter in galactic scaling relations and the stochasticity when sampling clusters from the ICMF and when assigning clusters to simulation particles. To study how much the resulting GC systems are influenced by the model randomness, we rerun the model 25 times on each dwarf galaxy with different random seeds. This allows us to present most results in terms of the median values and interquartile (25\%--75\%) ranges.

\section{Globular cluster systems of dwarf galaxies}
\label{sec:number_of_globular_clusters}

One of the most fundamental properties of observed GC systems in dwarf galaxies is the number of GCs. Since a large fraction of dwarf galaxies do not presently host any GCs, we divide the sample into two categories: galaxies with GCs and without GCs, and analyse them separately. We compare the model results with observations including the ELVES survey of LV GCs, the GC systems in the MW and MW/M31 satellites, and the catalogues of GC systems from \citet{harris_catalog_2013,harris_galactic_2017} and \citet{forbes_extending_2018}.

\subsection{Observational data in the Local Volume}
\label{sec:obs_data}

We compare the predictions of our model with the observational data from \citet{carlsten_elves_2022}. These authors analysed GC systems in the LV galaxies from the ELVES survey. This survey reviews satellite galaxies inside 300 projected kpc of luminous host galaxies ($M_{\rm V}<-22.1$) out to 12 Mpc of Earth. They investigated GC systems in a sample of 140 confirmed early-type dwarf satellite galaxies with stellar mass between $10^{5.5}$ and $10^{8.5}\Msun$ associated with 23 LV hosts. 

\citet{carlsten_elves_2022} obtained GC catalogues by identifying point sources in the surroundings of each dwarf galaxy. To exclude red sources that are unlikely to be GCs, they applied a color selection ${\rm g-r}\in[0.1,0.9]$ for dwarfs with $\rm g/r$ imaging or ${\rm g-i}\in[0.2,1.1]$ for dwarfs with $\rm g/i$ imaging. Additionally, they also applied a magnitude cut $M_{\rm g}\in(-9.5,-5.5)$. They determined the total number of GCs by counting GCs within twice of the dwarf's effective radius ($2r_{\rm e}$) and corrected the value for the incompleteness of faint GCs, GCs outside $2r_{\rm e}$, and the subtraction of background sources. They also applied an alternate likelihood method taking into account the magnitude and spatial distribution of candidate GCs. The magnitude distribution is modeled by a two-parameter Gaussian distribution, and the spatial distribution by a Plummer profile with a single parameter: $r_{\rm e}$. This method models the dwarf galaxies as a mixture of systems without GCs and with non-zero GCs. They parameterize the number of GCs as a two-parameter power-law function of the stellar mass of host galaxy, and the fraction of dwarfs with non-zero GCs as a monotonically increasing function of stellar mass characterized by values at 5 reference stellar masses. These accumulate to a total of 10 free parameters. The posterior distributions of the 10 parameters are obtained via Markov Chain Monte Carlo.

\subsection{Occupation fraction}
\label{sec:occupation_fraction}

A measure of stellar mass of dwarf satellite galaxies can be more easily obtained from observations than the total dynamical mass, and therefore it is beneficial to investigate how the properties of their GC systems scale with the stellar mass. On the other hand, our model is based on the halo mass, and the information about stellar mass comes only from applying the SMHM relation \citep{behroozi_average_2013}. Note that this relation is poorly constrained at the low-mass end, where the observed scatter is large and many physical processes that can introduce additional systematic bias are not considered. For example, applying this relation at $z=0$ may underestimate the actual stellar mass for satellite galaxies because of tidal truncation by the host galaxy. This truncation is likely to strip a higher fraction of halo mass than stellar mass, because stars are more compactly distributed even in satellite galaxies. Therefore, $\Mstar^{\rm z=0}$ is likely a lower limit on the actual stellar mass of the satellite. An opposite limit can be obtained by assuming that no stellar mass is stripped from the satellite. Then we could use the historical maximum value $\Mstar^{\rm max}$, which is the stellar mass resulting from applying the SMHM relation at the time of its peak value (typically, around the time of accretion onto the host). The true value of the satellite's stellar mass must lie between these two limits. Therefore, we treat the stellar mass of a model galaxy $i$ as obeying a continuous distribution between $M_{\star,i}^{\rm z=0}$ and $M_{\star,i}^{\rm max}$.

We employ two distribution functions for $\Mstar$. The first option is a simple flat function in the base-10 logarithmic space, i.e., 
\begin{equation}
    p_{\rm flat}(\log \Mstar) =
	\left\{ 
	\begin{array}{l}
		{\rm constant},\ {\rm if}\ \Mstar^{\rm z=0} \leq \Mstar < \Mstar^{\rm max}\\
		0,\ \mathrm{otherwise}.
	\end{array}
	\right.
\end{equation}
For visual clarity, with drop the $10$ subscript in $\log_{10}\Mstar$ for this expression and hereafter. This function assumes that the probability density for $\log \Mstar$ being any value within that range is the same. 

Alternatively, we may expect that the actual stellar mass is closer to $\Mstar^{\rm max}$, since stars are more concentrated towards the satellite center and less likely to be tidally stripped than the dark matter. To account for this, we introduce an alternate distribution which is a linear function in the logarithmic space:
\begin{equation}
    p_{\rm lin}(\log \Mstar)\propto 
	\left\{ 
	\begin{array}{l}
		\log \frac{\Mstar}{\Mstar^{\rm z=0}},\ {\rm if}\ \Mstar^{\rm z=0} \leq \Mstar < \Mstar^{\rm max}\\
		0,\ \mathrm{otherwise}.
	\end{array}
	\right.
\end{equation}
This function places more emphasis near $\Mstar^{\rm max}$. We will use both priors to investigate how GC numbers depend on stellar mass, and treat the difference in the results as systematic uncertainty associated with measuring $\Mstar$.

For example, when calculating the fraction of galaxies hosting at least one GC (`GC occupation fraction' $f_{\rm occ}$) as a function of $\Mstar$, we take the weighted average value in bins using a kernel smoothing method:
\begin{equation}
    f_{{\rm occ},j} = \frac{\sum_i^{\text{non-zero GCs}} \int_{M_{j}}^{M_{j+1}} p_i(\log \Mstar)\, d\log \Mstar}
    {\sum_i^\text{all galaxies} \int_{M_{j}}^{M_{j+1}} p_i(\log \Mstar)\, d\log \Mstar}.
    \label{eq:weighted_value}
\end{equation}
The summation in the denominator is over all galaxies, while the summation in the numerator is over galaxies with non-zero GCs. We define the term `non-zero GCs' as galaxies that contain at least one cluster above a certain lower mass limit $M_{\rm low}$. We set a default value $M_{\rm low}=3\times10^4\Msun$ to mimic the $M_{\rm g}<-5.5$ magnitude cut employed in the LV observations. Since $f_{\rm occ}$ can depend significantly on $M_{\rm low}$, we compare the results for different choices of $M_{\rm low}$ in Sec.~\ref{sec:comparing_settings}.

We also require the GCs to be located inside a certain radius from the galaxy center. Here, we set this radius to be an estimate of the tidal radius in an isothermal density potential, $r_{\rm tid} = d_{\rm host} (M_{\rm sat}/2M_{\rm host})^{1/3}$. Even though the actual tidal radius may not be used when identifying satellite GCs in observations, it is a physically meaningful proxy to use in the model.

The kernel function $p_i(\log \Mstar)$ in equation~(\ref{eq:weighted_value}) is either $p_{\rm flat}$ or $p_{\rm lin}$: the distribution function of stellar mass for the $i$-th galaxy. We obtain 25 values for $j$-th mass bin from the 25 random model realizations. We present the final result as the median of the 25 values, as well as the scatter represented by the interquartile range of the 25 values.

\begin{figure}
    \includegraphics[width=\linewidth]{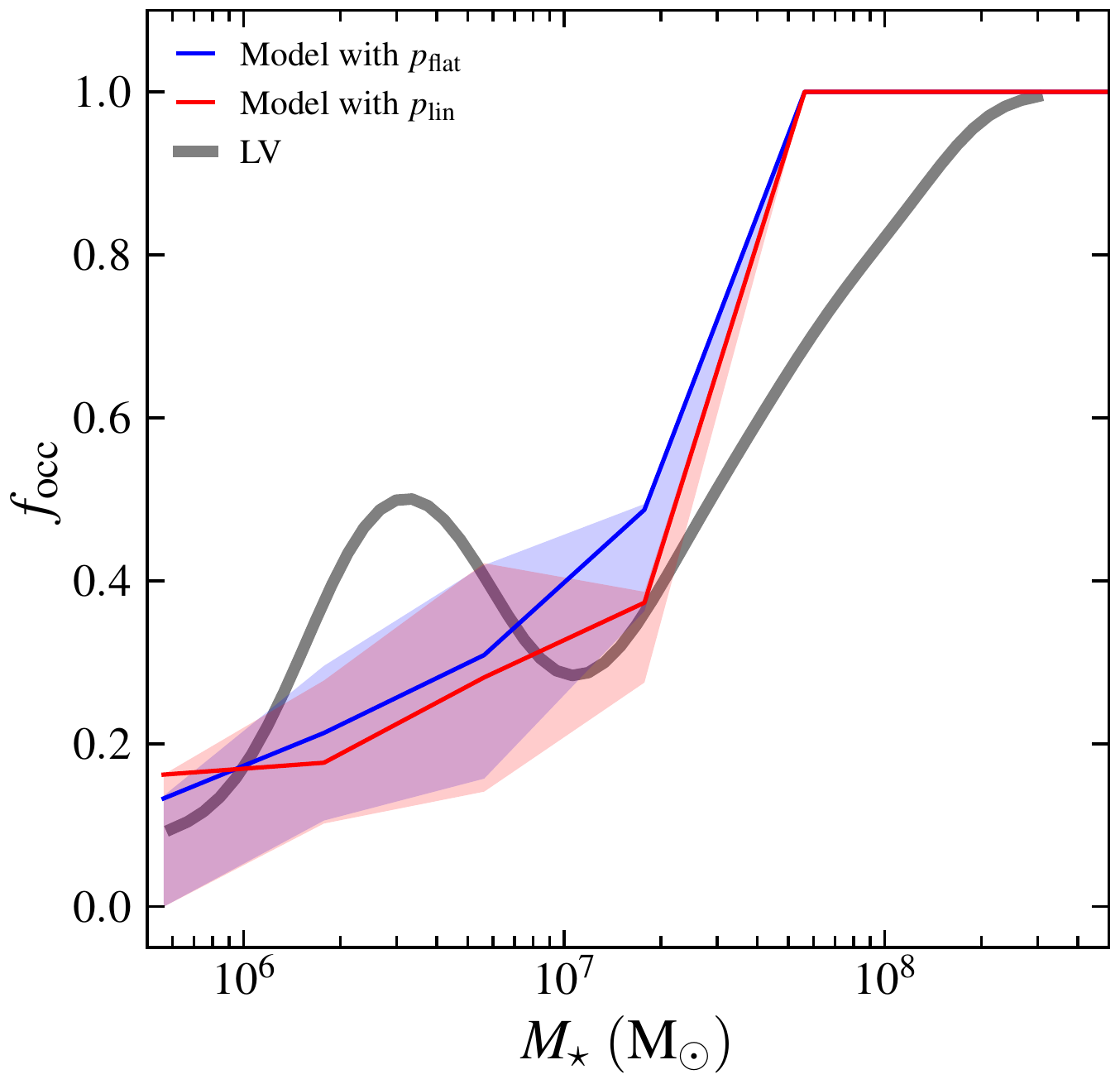}
    \vspace{-4mm}
    \caption{GC occupation fraction $f_{\rm occ}$ relation as a function of stellar mass of host galaxy $\Mstar$. The $f_{\rm occ}$--$\Mstar$ relations of model galaxies are shown as the cyan and magenta curves, employing the flat and linear distribution functions of $\Mstar$, respectively. The gray curve represents observed relation from the LV \citep{carlsten_elves_2022} with Gaussian kernel smoothing. See the main text for a detailed description of how we obtain these curves.}
    \label{fig:occupation_frac}
\end{figure}

In Fig.~\ref{fig:occupation_frac}, we show the occupation fraction as a function of stellar mass of the satellite galaxy. We note that $p_{\rm flat}$ and $p_{\rm lin}$ produce similar $f_{\rm occ}$--$\Mstar$ relations, although the relation from $p_{\rm lin}$ is shifted slightly to the high-mass end. For both distribution functions, the occupation fraction is almost 1 for $\Mstar\gtrsim5\times10^7\Msun$ but drops to less than 0.2 for $\Mstar\lesssim10^6\Msun$. At $\Mstar\simeq2\times10^7\Msun$, the occupation fraction is approximately 0.5.

The observational $f_{\rm occ}$--$\Mstar$ relation is also shown in Fig.~\ref{fig:occupation_frac} for comparison. The observed relation is computed via the kernel smoothing method with a Gaussian kernel. The expression for the Gaussian kernel smoothing method is also given by equation~(\ref{eq:weighted_value}), but replacing $p_i$ with a Gaussian function. Since the uncertainty of stellar mass in the ELVES data is $\gtrsim0.1$ dex \citep{carlsten_structures_2021}, we set the standard deviation of the Gaussian function to be $\sigma_{\log M}=0.2$ dex. The number of GCs provided by the ELVES data ($N_{\rm obs}$) is not necessarily an integer (or even positive) due to background subtraction. Therefore, we round $N_{\rm obs}$ to the nearest integer and define $f_{\rm occ}$ as the fraction of galaxies hosting at least one GC (which can be equivalently defined as the fraction of galaxies with $N_{\rm obs}>0.5$).

This method is different from the likelihood method employed in \citet{carlsten_elves_2022}, who enforced $f_{\rm occ}$ to be a monotonically increasing function of $\Mstar$. In contrast, we find the observational relation to be non-monotonic as $f_{\rm occ}$ has a spike at $\Mstar\simeq 3\times10^6\Msun$. We do not investigate the origin of this spike in depth, as it is not the main focus of this work. Despite the spike, the observed occupation fraction is around 1 for $\Mstar\gtrsim3\times10^8\Msun$ and also drops to less than 0.2 for $\Mstar\lesssim10^6\Msun$. At $\Mstar\simeq3\times10^7\Msun$, the occupation fraction is about 0.5. The model relation generally agrees with the observations except that the transition from $f_{\rm occ}=1$ to $0$ is steeper than the observed relation. It is surprising that our model shows good agreement with the observed satellite galaxies even if the model parameters are only calibrated for the 3 MW-like central galaxies with the MW GCs, suggesting that GC formation and evolution in satellite galaxies can be described by the same physical processes as the central galaxy.

\begin{figure}
    \includegraphics[width=\linewidth]{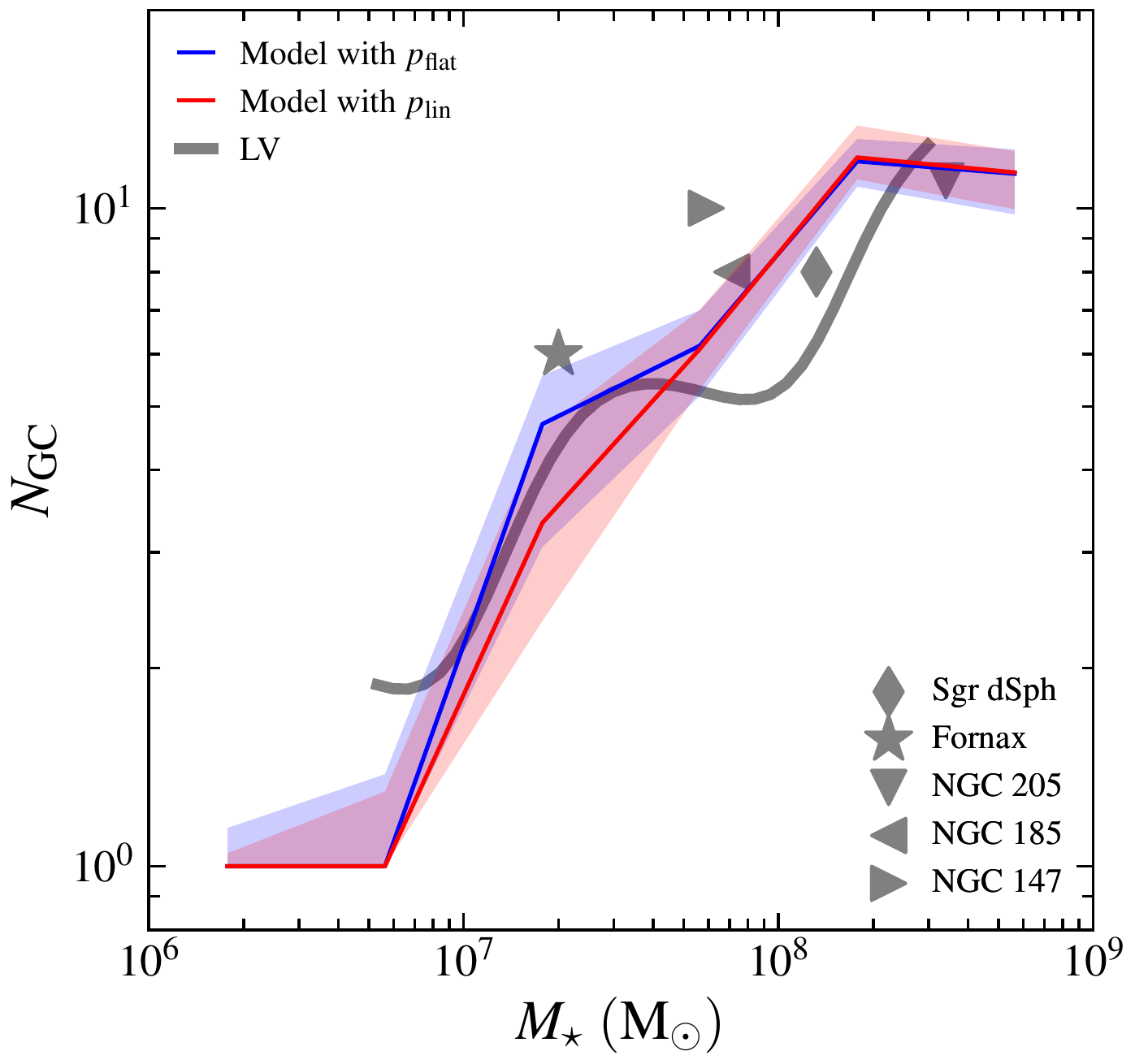}
    \vspace{-4mm}
    \caption{Number of GCs $\Ngc$ as a function of stellar mass of host galaxy $\Mstar$. We plot the $\Ngc$--$\Mstar$ relation from Gaussian kernel smoothing for the LV dwarf galaxies \citep{carlsten_elves_2022} as the gray curve. Satellites of the MW and M31 are shown as gray symbols.}
    \label{fig:Ngcs_Ms}
\end{figure}

\subsection{Number of globular clusters}

For the dwarf galaxies with non-zero GCs, we further investigate the relation between $\Ngc$ and stellar mass of the galaxy. We calculate this relation in a similar fashion to the occupation fraction. After splitting $\Mstar$ into bins, we compute the weighted average of the $j$-th bin using a kernel smoothing method:
\begin{equation}
    N_{{\rm GC},j} = \frac{\sum_i^\text{non-zero GCs} \int_{M_{j}}^{M_{j+1}} N_{{\rm GC},i} \, p_i(\log \Mstar) \, d\log \Mstar}
    {\sum_i^\text{non-zero GCs} \int_{M_{j}}^{M_{j+1}} p_i(\log \Mstar) \, d\log \Mstar}
    \label{eq:weighted_value_n}
\end{equation}
where $N_{{\rm GC},i}$ stands for the number of clusters of the $i$-th galaxy. Again, we define $\Ngc$ to be the number of clusters with mass above $M_{\rm low}$, which is set by default to $3\times10^4\Msun$ to mimic the observed magnitude cut. Here, both summations are over galaxies with non-zero GCs.

In Fig.~\ref{fig:Ngcs_Ms}, we show the $\Ngc$--$\Mstar$ relation for model galaxies and compare it with the observed relation for the LV dwarf galaxies with Gaussian kernel smoothing similar to the calculation of the occupation fraction. In addition to the LV dwarf galaxies, we also compare our results with several satellites of the MW and M31 systems, including the Sagittarius dwarf spheroidal galaxy \citep[Sgr dSph,][]{law_assessing_2010}, Fornax dSph galaxy \citep{pace_spectroscopic_2021}, NGC 205 \citep{da_costa_globular_1988}, NGC 185 \citep{veljanoski_newly_2013}, and NGC 147 \citep{veljanoski_newly_2013}. The observed number of GCs is very uncertain for some satellites. For example, \citet{law_assessing_2010} suggested that Sgr dSph hosts 8 GCs, whereas \citet{minniti_discovery_2021} almost tripled this number to 23. However, this does not affect qualitatively the comparison with observations since we are interested in a general trend of the $\Ngc$--$\Mstar$ relation rather than reproducing the exact number of GCs in a particular observed galaxy.

We find that the $\Ngc$--$\Mstar$ relations obtained with $p_{\rm flat}$ and $p_{\rm lin}$ are consistent with each other. Both relations show that the number of GCs is around 10 at $\Mstar=10^8\Msun$ and decreases monotonically to less than 2 at $\Mstar\lesssim10^7\Msun$ (note that $\Ngc$ is always greater than 0 since we only analyse here galaxies with non-zero GCs). The modeled $\Ngc$ continue to drop to around 1 at $\Mstar\lesssim5\times10^6\Msun$. We do not plot the observed relation below $\Mstar=5\times10^6\Msun$ because $\Ngc$ from the ELVES survey may be influenced by numerical bias at small $\Mstar$: the numbers of GCs in LV dwarf galaxies provided by \citet{carlsten_elves_2022} are corrected for GCs outside 2$r_{\rm e}$ annulus by dividing the GC counts by a factor of $0.646$. Additionally, they apply a small factor to correct for GCs below the detection limit. Such corrections boost GC counts by a factor of around 1.6, leading to numerical bias in the low-mass end where the lowest non-zero GC count is 1. Therefore, $\Ngc$ of LV galaxies drops to a minimum value $\gtrsim 1.6$ instead of 1 at $\Mstar\lesssim5\times10^6\Msun$. It is meaningless to compare the model and observed relations in this regime. Despite this region, our model predicts the $\Ngc$--$\Mstar$ relation in consistency with observations from $5\times10^6$ to $3\times10^9\Msun$. We emphasize that the model is only calibrated for the MW-like galaxies with $\Mstar>10^{10}\Msun$, the good agreement at such a low-mass range is not a trivial outcome of tuning the model parameters. Instead, it implies that physical processes controlling GC formation and evolution may be universal for both central and satellite galaxies. We also note that the model relations have significant scatter at all masses, which increases the uncertainty when trying to apply the $\Ngc$--$\Mstar$ relation to estimate the stellar mass and number of GCs of a dwarf galaxy.

\subsection{Scaling with stellar mass}
\label{sec:scaling_with_stellar_mass}

\begin{figure*}
\includegraphics[width=0.49\linewidth]{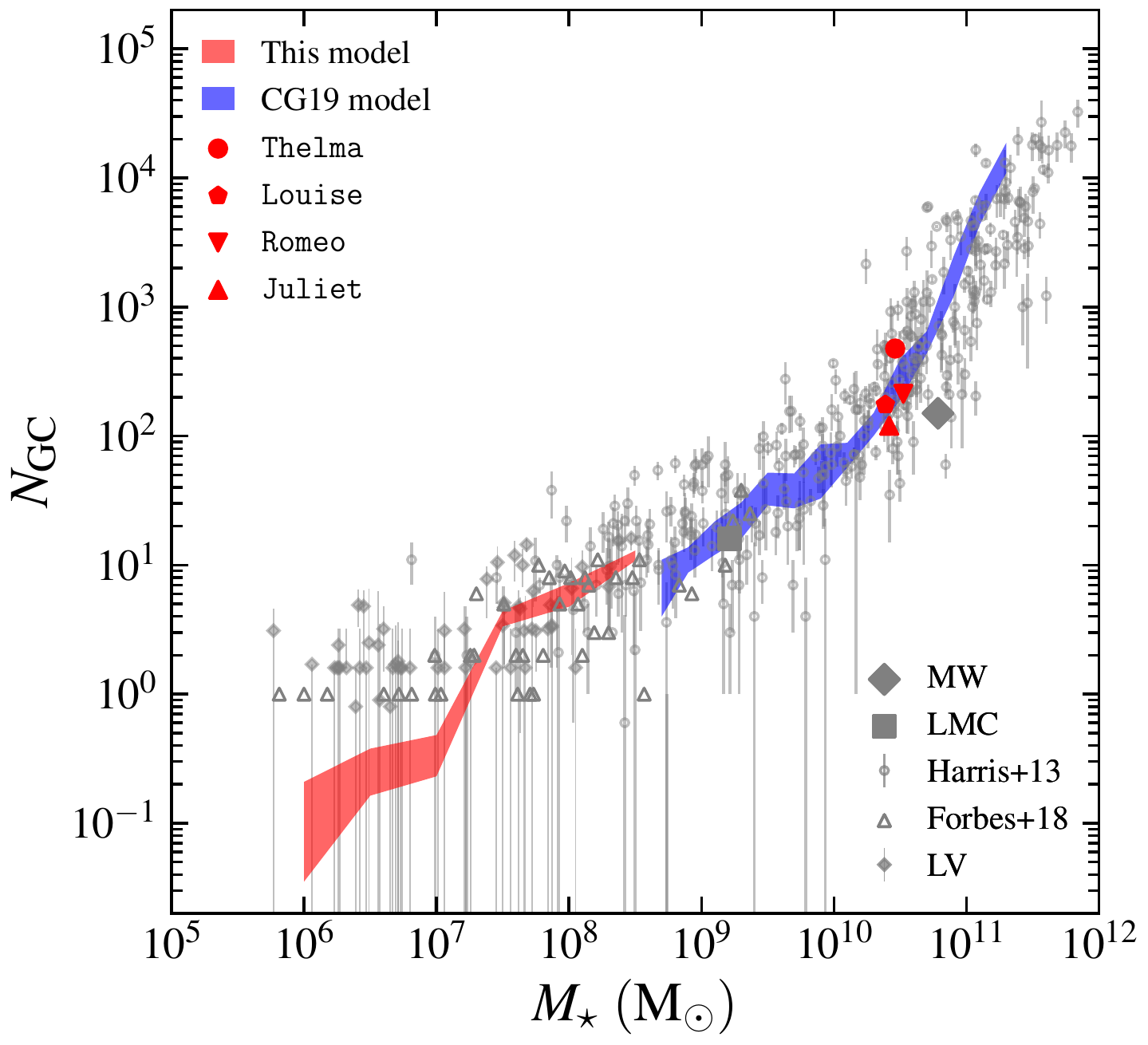}
\includegraphics[width=0.49\linewidth]{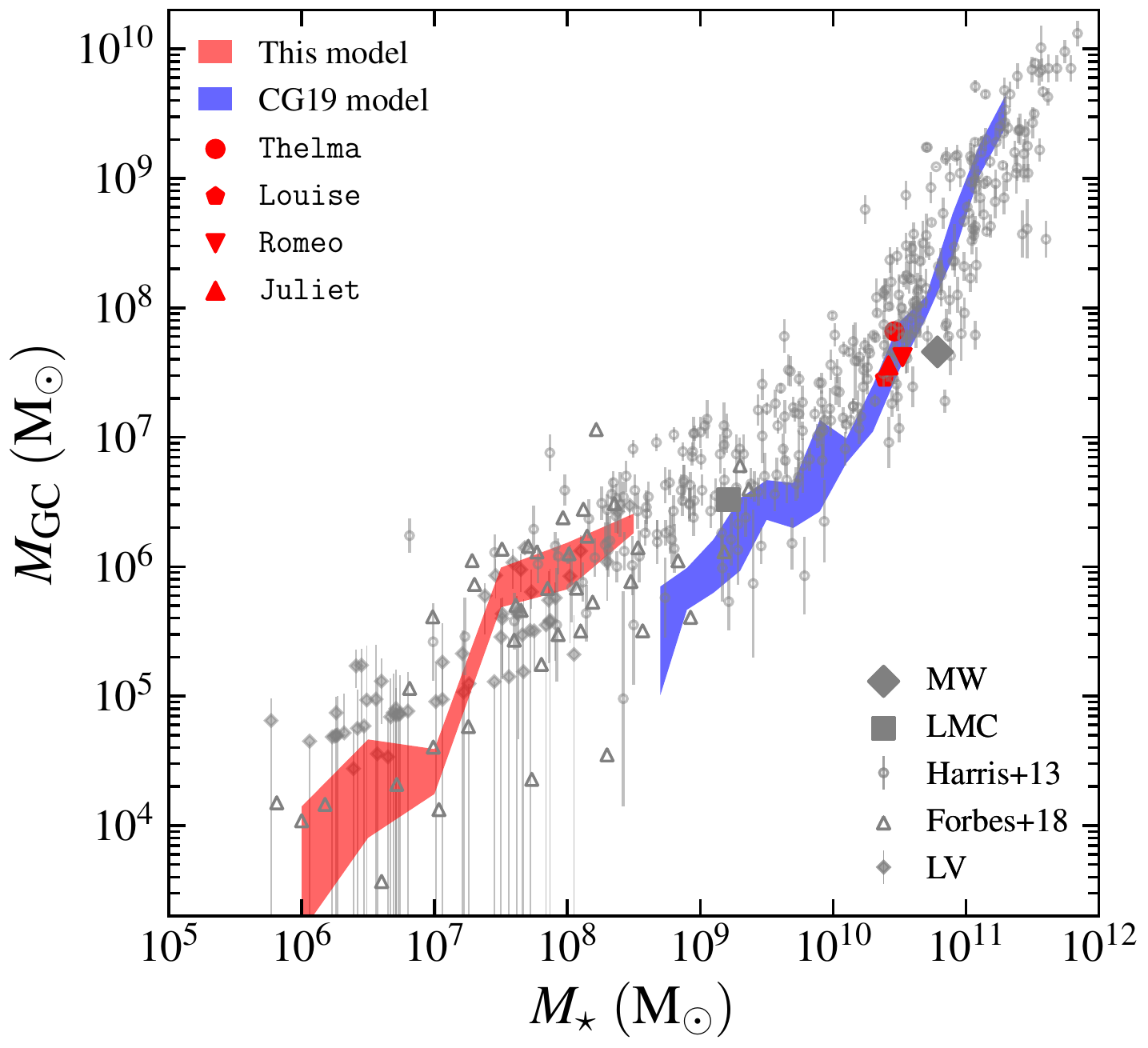}
  \vspace{-1mm}
  \caption{Number of GCs (\textit{left panel}) and total mass of GCs (\textit{right panel}) vs the stellar mass of host galaxy. Red shaded region shows the interquartile range of the present model, blue shaded region shows the interquartile range of the model from \citet{choksi_origins_2019}. We plot the observational data from \citep{harris_catalog_2013} as gray circles with errorbars, the data from \citet{forbes_extending_2018} as open triangles, and the data from the LV \citep{carlsten_elves_2022} as diamonds with errorbars. We also plot the four main galaxies in the simulations as the red symbols, with the MW and LMC shown as the gray diamond and square for comparison.}
  \label{fig:Ngcs_Ms_wide}
\end{figure*}

In this section, we extend our comparison with observations to a broader stellar mass range. In Fig.~\ref{fig:Ngcs_Ms_wide}, we show the $\Ngc$/$\Mgc$--$\Mstar$ relations, where $\Mgc$ stands for the total mass of the GC system. For clarity we show only relations using the $p_{\rm lin}$ distribution function, since the two distribution functions $p_{\rm flat}$ and $p_{\rm lin}$ give consistent results. Different from the above analysis, we also include dwarfs with zero GCs in the calculation of $\Ngc$. That is, the summations in equation~(\ref{eq:weighted_value_n}) are over all galaxies instead of galaxies with non-zero GCs only. This setting allows $\Ngc$ to drop below 1 for the lowest-mass galaxies. The model $\Ngc$ behaves similarly to Fig.~\ref{fig:Ngcs_Ms} for $\Mstar>10^7\Msun$, where most galaxies have at least one GC, and continues to drop to $\sim0.1$ at $\Mstar=10^6\Msun$ since a large fraction of dwarf galaxies do not actually host any cluster \citep[see,][and relevant discussion in Sec.~\ref{sec:occupation_fraction}]{carlsten_elves_2022}. We notice that $\Ngc$ drops significantly at $\Mstar=(1-3)\times10^7\Msun$. However, we have no evidence that such an abrupt decline is physically real since only 2 galaxies lie in this range. The poor statistics in this narrow range is unreliable. Therefore, we only focus on the scaling relations across a wider mass range ($\gtrsim$1 order of magnitude) which includes more galaxies ($\sim$10). The $\Mgc$--$\Mstar$ relation has a similar trend. Starting from $\Mgc\simeq 10^6\Msun$ at $\Mstar=10^8\Msun$, $\Mgc$ drops to $\sim10^4\Msun$ at $\Mstar=10^6\Msun$, i.e., $\sim1\%$ of the total stellar mass resides in surviving GCs. For comparison, $5-40\%$ of total stellar mass was originally formed in GCs (with initial mass $>10^4\Msun$). At $z\gtrsim 5$, this fraction approaches (even slightly exceeds) $100\%$, indicating that star formation is dominated by cluster formation at early epochs. The slight excess of cluster formation rate may be due to the potentially underestimated star formation rate by the \citet{behroozi_average_2013} SMHM, which is poorly constrained at the low mass end and at high redshift. The subsequent tidal disruption significantly reduces the fraction of stars in clusters to its present-day value $\sim1\%$.

For comparison, we plot the observed GC systems in the MW \citep[stellar mass from][]{licquia_improved_2015} and Large Magellanic Cloud (LMC) as well as the catalogues of GC systems from the LV survey, the catalogues by \citet{harris_catalog_2013} and \citet{forbes_extending_2018}. The latter catalogue focuses on GC systems in the LG dwarf galaxies down to stellar mass $\lesssim10^6\Msun$. Note that the \citet{harris_catalog_2013} catalogue does not list $\Mstar$ directly. Instead, it provides K-band magnitude $M_{\rm K}$. We convert $M_{\rm K}$ to $\Mstar$ using a fixed stellar mass-to-light ratio $\log(\Mstar/L_{\rm K})=-0.3$ \citep[estimated from Fig.~20 in][]{bell_optical_2003}. Moreover, the LV data do not provide the total mass of GC systems. To estimate $\Mgc$ from $\Ngc$, we fit the mean GC mass in \citet{forbes_extending_2018} as a power-law function of $\Mstar$: $\log\overline{M}=2.3+0.35\log\Mstar$, and compute the total GC mass as $\Mgc=\Ngc\overline{M}$. 

We also overplot in Fig.~\ref{fig:Ngcs_Ms} the $\Ngc$/$\Mgc$--$\Mstar$ relations from a previous version of our model \citep{choksi_formation_2019}, which successfully matches the observational trend for $10^{9.5}<\Mstar< 10^{11}\Msun$. However, that model deviates from observations at $\Mstar\lesssim10^{9.5}\Msun$. Compared to the scaling of $\Ngc$, the deviation is more significant for $\Mgc$. This is partly because the previous model has inefficient disruption of low-mass clusters (the $y=2/3$ case in Sec.~\ref{sec:cluster_evolution}). Therefore, such a prescription predicts a GC mass function peaked at lower mass, and thus tends to underestimate the mean mass of surviving clusters, as shown in Fig.~\ref{fig:mass_function}. Our updated model attempts to solve the deviation by following the formation of less massive clusters (down to $10^4\Msun$) in low-mass galaxies (down to $\Mh=10^8\Msun$). In addition, the current model applies a more realistic tidal disruption prescription taking into account the local environment of clusters. The resulting $\Ngc$/$\Mgc$ of the new model can match the observed relations at a mass range where most galaxies have non-zero GCs, $\Mstar\gtrsim3\times10^7\Msun$. Below this mass, the model $\Ngc$ continues to drop below 1, while the observational data on dwarfs from \citet{forbes_extending_2018} only consider galaxies with non-zero clusters, leading to the observed $\Ngc$--$\Mh$ relation bending upwards at the low-mass end. It is therefore meaningless to compare the $\Ngc$/$\Mgc$--$\Mstar$ relations at $\Mstar\lesssim3\times10^7\Msun$. We still need better observations of dwarf galaxies to further test the validity of our model.

\subsection{Scaling with halo mass}
\label{sec:scaling_with_halo_mass}

\begin{figure*}
\includegraphics[width=0.49\linewidth]{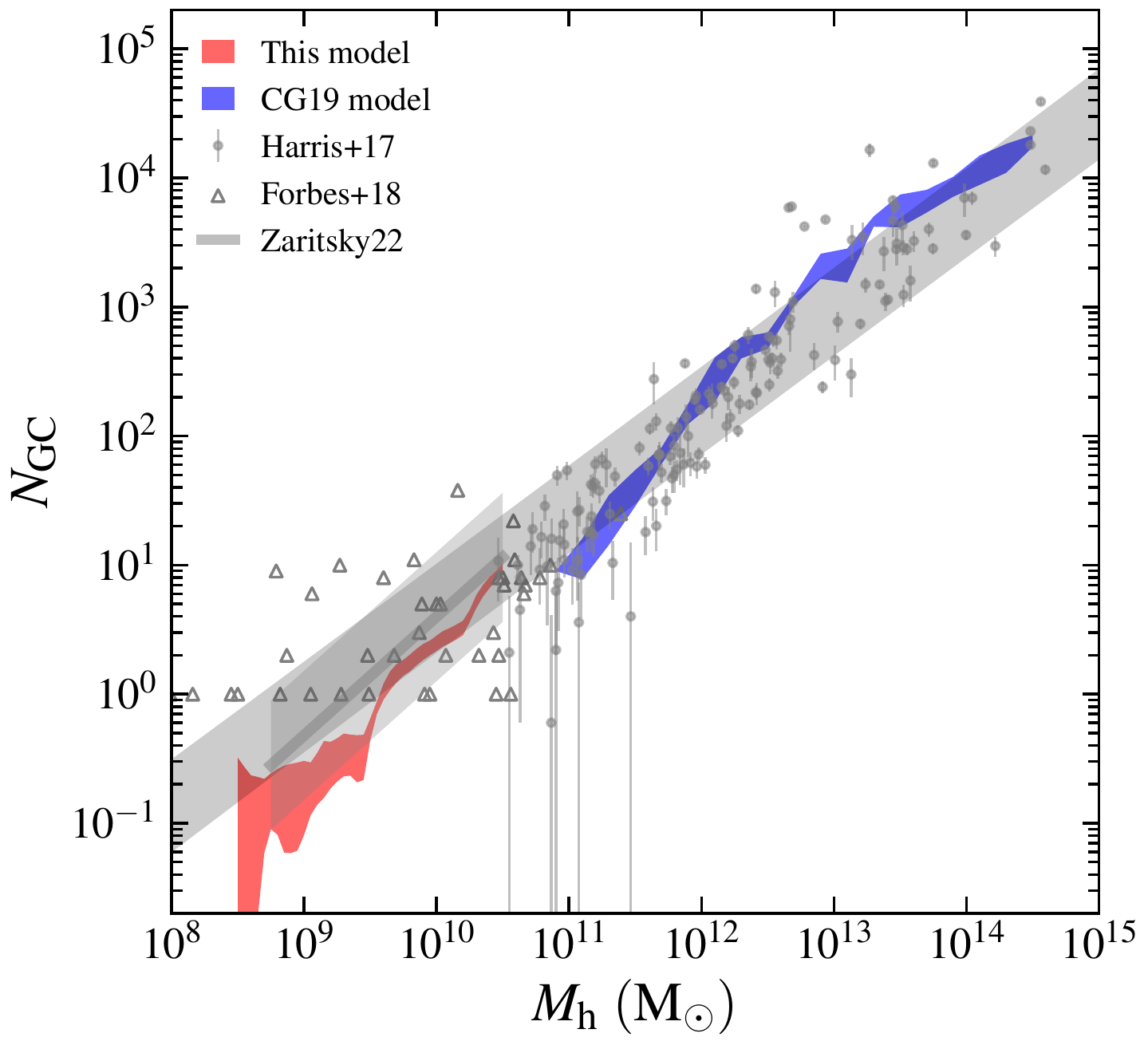}
\includegraphics[width=0.49\linewidth]{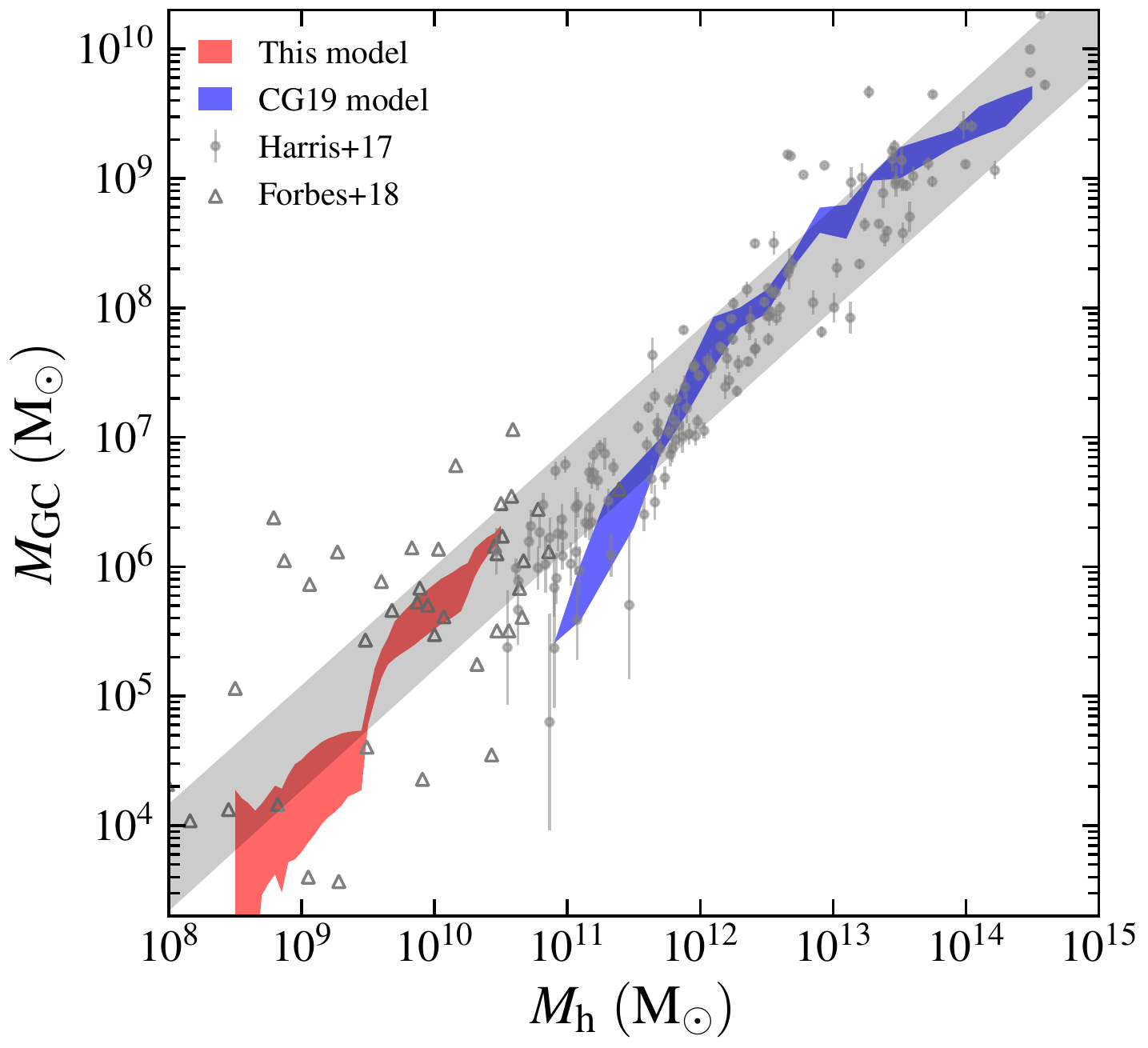}
  \vspace{-1mm}
  \caption{Number of GCs (\textit{left panel}) and total mass of GCs (\textit{right panel}) vs the total mass of host galaxy. Red shaded region shows the interquartile range of the present model, blue shaded region shows the interquartile range of the model from \citet{choksi_origins_2019}. We plot the observational data from \citep{harris_galactic_2017} as gray circles with errorbars, and data from \citet{forbes_extending_2018} as open triangles. The long gray region shows the jointly fitted power-law relation of the two observational datasets with intrinsic scatter. The power-law dependence from \citet[][$\Ngc\propto\Mh^{0.92}$]{zaritsky_revisiting_2022} is shown as the gray line in the \textit{left panel}, with the estimated scatter $0.5$ dex plotted as the short gray region. The gray shaded region shows the power-law fit of the observational data, see the main text for more details.}
  \label{fig:Ngcs_Mh}
\end{figure*}

After investigating the dependence of $\Ngc$/$\Mgc$ on stellar mass and showing that it is consistent with observations, we can turn to the dependence on the satellite halo mass. In the left panel of Fig.~\ref{fig:Ngcs_Mh}, we present the model $\Ngc$--$M_{\rm h}$ relation via the kernel smoothing method with an \citet{epanechnikov_non-parametric_1969} kernel: $K(u) = 0.75(1-u^2)$ for $|u|\leq1$ and $K(u) = 0$ otherwise. The bandwidth is set to 0.5 dex. We find that different bandwidths from 0.4 to 1 dex do not alter the relation significantly (however, a bandwidth $<$ 0.4 dex is insufficient to cover all gaps between neighbouring data points).

We show the observational data from \citet{harris_galactic_2017} and \citet{forbes_extending_2018} for comparison. The first catalogue covers galaxies with halo mass between $10^{9.5}-10^{14.5}\Msun$, and the second catalogue focuses on LG dwarf galaxies with $\Mh=10^8-10^{11}\Msun$. We fit the two datasets jointly with a power-law:
\begin{equation}
    \log \Ngc = a + b\log M_{\rm h12} + \epsilon
\end{equation}
where $M_{\rm h12}$ is the halo mass $\Mh$ in unit of $10^{12}\Msun$. The intrinsic scatter is represented by a random variable $\epsilon$ following a Gaussian distribution ${\cal N}(0, \sigma_{\rm int})$. We perform the fit by maximizing the likelihood:
\begin{equation}
    {\cal L}\equiv\prod_i \frac{1}{\sigma_i\sqrt{2\pi}} \exp\left( -\frac{1}{2} \frac{\delta_i^2}{\sigma_{\log N,i}^2 + \sigma_{\rm int}^2} \right),
\end{equation}
where $\delta_i=\log N_{{\rm GC},i}-a-b\log M_{{\rm h12},i}$ is the vertical deviation, with the subscript $i$ denoting the $i$-th data point. In addition, $\sigma_{\log N,i}$ is set to the observed uncertainty of $\log N_{{\rm GC},i}$ if provided or $0.3$ dex otherwise. We apply bootstrap resampling $1000$ times until all fitting parameters converge to estimate the mean values and standard deviations of $a$, $b$, and $\sigma_{\rm int}$. Maximizing the likelihood ${\cal L}$ yields
\begin{equation}
    \log \Ngc = (2.20\pm0.03) + (0.77\pm0.03)\log M_{\rm h12}
    \label{eq:simple_fit_model}
\end{equation}
with an intrinsic scatter $\sigma_{\rm int}=(0.34\pm0.03)$ dex.

Again, we show in Fig.~\ref{fig:Ngcs_Mh} the $\Ngc$--$\Mh$ relation from \cite{choksi_formation_2019}. This model successfully matches the observational trend for $10^{12}<M_{\rm h}< 10^{14}\Msun$. Like the $\Ngc$--$\Mstar$ relation, this model underestimates $\Ngc$ at the low-mass end $\Mh\sim10^{11}\Msun$. In the contrast, although the new model still falls systematically below the observed $\Ngc$ in the dwarf galaxy range we study here, it is within a factor of $\sim2$ of the average trend. Moreover, the observational trend may be biased upwards as the data of dwarfs from \citet{forbes_extending_2018} only consider galaxies with non-zero clusters, and they measured halo mass systematically below the derived values from the majority of SMHM relations. A more appropriate comparison in the lowest-mass end is with \citet{zaritsky_revisiting_2022}, who revisited the observational data by \citet{forbes_globular_2020} and \citet{carlsten_elves_2022}. These two datasets both provide information about galaxies hosting zero GCs. \citet{forbes_globular_2020} studied the GC systems in the  Coma cluster ultra-diffuse galaxies (UDGs). \citet{zaritsky_revisiting_2022} reconciled the number of GCs in \citet{forbes_globular_2020} by multiplying a factor of $0.27$, taking into account a more precise constraints on GC luminosity function and radial distribution \citep{saifollahi_implications_2022}. \citet{zaritsky_revisiting_2022} also derived total mass of galaxies from the two existing catalogues using an extension of the fundamental plane formalism. This method uses empirically calibrated relations to estimate the mass-to-light ratio within the half-light radius and the enclosed mass. The author then fit an NFW profile to the DM components inside the half-light radius to obtain the total mass of the galaxy. The author discovered a near-linear $\Ngc$--$\Mh$ relation for both \citet{forbes_globular_2020} and \citet{carlsten_elves_2022} samples: $\Ngc\propto\Mh^{0.92\pm0.08}$. By plotting this relation in Fig.~\ref{fig:Ngcs_Mh}, we find that it is consistent with the observed power-law relation from \citet{harris_galactic_2017} and \citet{forbes_extending_2018}. The model relation shows a similar slope but lies $\sim0.3$ dex below. Considering the large scatter of the $\Ngc$--$\Mh$ relation found by \citet[][]{zaritsky_revisiting_2022}, at least $\sim 0.5$ dex, our model still makes predictions consistent with observations.

However, it is worth noting that \citet{forbes_globular_2020} studied GCs in UDGs instead of satellite galaxies as analysed in this work and in \citet{carlsten_elves_2022}. These UDGs normally have more GCs than the dwarfs of the same stellar mass, indicating that these galaxies have greater total mass than the predictions from typical SMHM. Although \citet{zaritsky_revisiting_2022} suggested that the $\Ngc$--$\Mh$ relation from UDGs is consistent with the relation from the LV satellites, the two categories of galaxies may follow different formation scenarios and are not directly comparable. 

In addition to the $\Ngc$--$\Mh$ relation, we also investigate the $\Mgc$--$\Mh$ relation from the model. This relation follows an interesting near-linear scaling across a broad mass range \citep{spitler_new_2009,georgiev_globular_2010,hudson_dark_2014,harris_dark_2015,forbes_extending_2018}. However, this scaling is only confirmed for galaxies with $M_{\rm h}>10^{10}\Msun$ since it is challenging to determine the halo mass of dwarf galaxies directly. With the observational data from \citet{forbes_extending_2018}, we attempt to extend the relation to low-mass galaxies. Similarly to the $\Ngc$--$\Mh$ relation, we fit the observational data from \citet{harris_galactic_2017} and \citet{forbes_extending_2018} jointly and obtain a power-law relation
\begin{equation}
    \log \Mgc = (7.45\pm0.03) + (0.93\pm0.03)\log M_{\rm h12}
\end{equation}
with an intrinsic scatter $\sigma_{\rm int}=(0.39\pm0.04)$ dex. The slope of $0.93\pm0.03$ is very close to unity, meaning that we can extend the near-linear relation down to $\Mh\sim 10^8\Msun$. In the right panel of Fig.~\ref{fig:Ngcs_Mh}, we compare the model $\Mgc$--$\Mh$ relation with this observational relation. Compared to the $\Ngc$--$\Mh$ relation, $\Mgc$ from the \cite{choksi_formation_2019} model deviates even more from the observations at $\Mh\sim 10^{11}\Msun$ because it underestimates the mean mass of surviving clusters, see the discussion in Sec.~\ref{sec:scaling_with_stellar_mass}. In contrast, the new model is in good agreement with observations as the model relation mostly overlaps the observed $\Mgc$ in dwarfs. It is remarkable that our model can match the observational relation down to $\Mh\sim 10^8\Msun$, extending this near-linear correlation to $\sim6$ orders of magnitude of halo mass even considering the complicated interplay of multiple non-linear processes in GC formation. 

However, we emphasize that the SMHM relation, which is widely used to convert observed stellar mass to halo mass, and simulated halo mass to stellar mass, is not well constrained at the low-mass end. This is due to the scarcity of independent measurements of halo mass for dwarf galaxies. Also, the scatter of SMHM may be underestimated in the \citet{behroozi_average_2013} relation, which assumes a constant scatter for all masses. More detailed observations are needed to better constrain the SMHM in the dwarf range and that may change our current knowledge of the $\Ngc$/$\Mgc$--$\Mstar$ relation.

\section{Detailed example: analogue of Fornax dSph galaxy}
\label{sec:detailed_example}

\begin{figure*}
    \includegraphics[width=\linewidth]{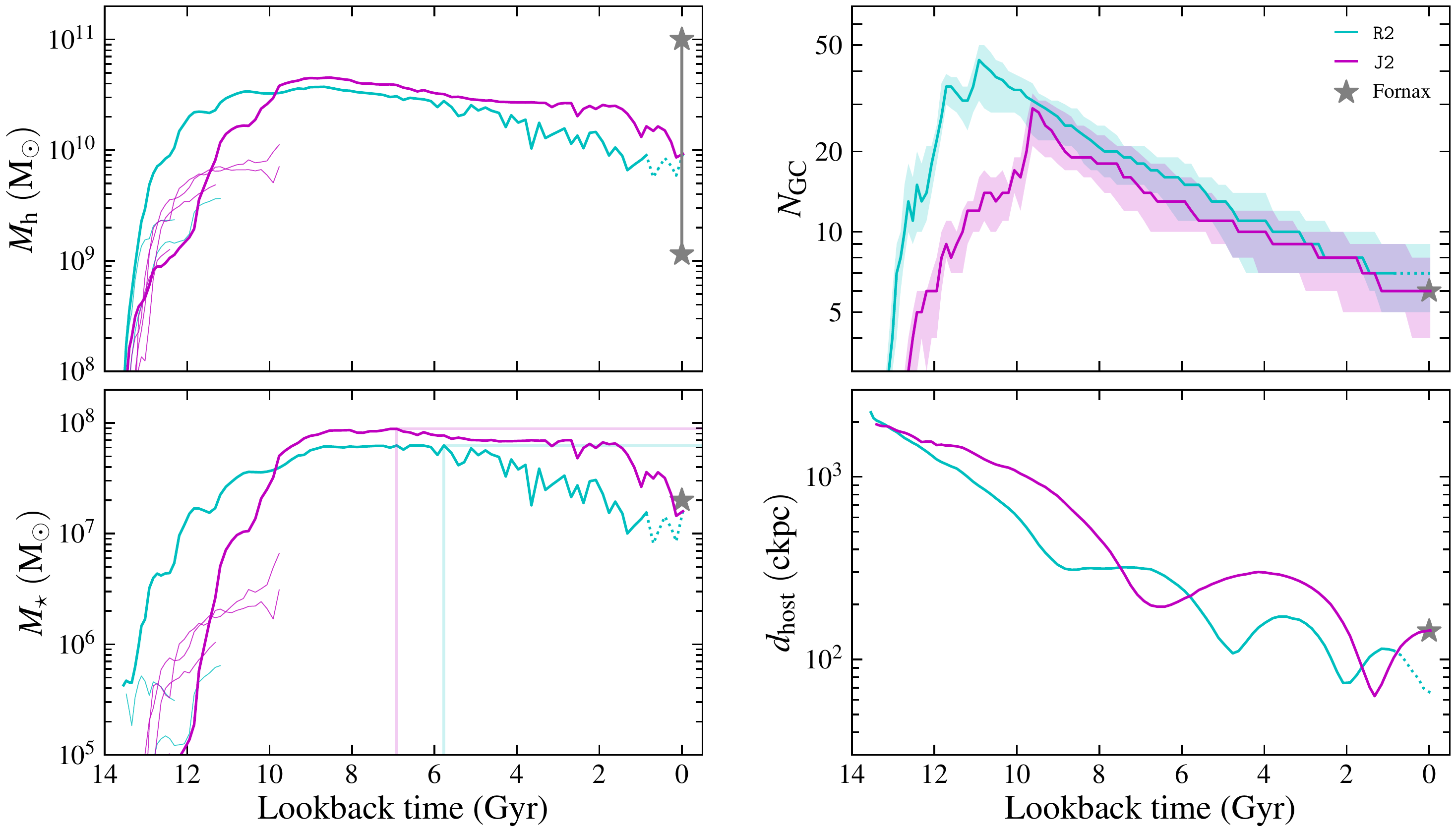}
    \vspace{-4mm}
    \caption{Time evolution of halo mass $M_{\rm h}$ (\textit{upper left}), stellar mass $\Mstar$ (\textit{lower left}), number of GCs $\Ngc$ (\textit{upper right}), and the distance to the host galaxy $d_{\rm host}$ (in comoving kpc, \textit{lower right}) of the two Fornax-like dwarf galaxies in the \texttt{R\&J} run. We plot them as solid curves until the epoch when we consider these two galaxies to be the most similar to the Fornax dSph. After that, we plot the curves in dotted style. We also plot with thin lines the mass histories of their own satellites that contribute at least one surviving cluster to the present-day GC population. We mark the historical maximum of $\Mstar$ as intersections of the vertical and horizontal lines in the \textit{lower left} panel. In the \textit{upper right} panel, The time evolution of $\Ngc$ is shown as solid curves with shaded regions, representing the median value and the interquartile range from the 25 random realizations. The Fornax dSph galaxy is represented by gray stars in each panel. Since different works predict vastly different halo masses for Fornax, ranging from $M_{\rm h}\sim10^9\Msun$ \citep{forbes_extending_2018} to $M_{\rm h}\sim10^{11}\Msun$ \citep[obtained from the SMHM relation by][]{danieli_elves_2022}, we show the two extreme halo masses in the \textit{upper left} panel for completeness.}
    \label{fig:evolv_plots}
\end{figure*}

In this section we show a detailed example of how GC systems in satellite galaxies evolve over cosmic time. We focus on the most massive satellite of the \texttt{Romeo} (or \texttt{R2}) and \texttt{Juliet} (or \texttt{J2}) galaxies in the \texttt{R\&J} simulation. These two satellites resemble the Fornax dSph galaxy in many aspects. In Fig.~\ref{fig:evolv_plots} we show the time evolution of the halo mass $M_{\rm h}$, stellar mass $\Mstar$, number of clusters $\Ngc$, and the distance to the host galaxy $d_{\rm host}$. \texttt{J2} is $140$ kpc away from the host galaxy at present, in close agreement with the Fornax dSph which is also about $140$ kpc away from the MW center. We compute the tidal radius of \texttt{J2} to be 24 kpc. The other satellite (\texttt{R2}) is closer ($\sim 70$ kpc) to the host galaxy as it is near the pericentre. However, the properties of this satellite agree better with those of Fornax dSph if we look back for 1 Gyr, when this satellite is near its apocenter about 110 kpc away from the host galaxy, with a corresponding tidal radius being 17 kpc. Therefore, we consider $t_{\rm lookback}=$ 1 Gyr as the `present-day' for \texttt{R2} and only focus on the properties at/before this epoch when referring to this galaxy. In addition, we find that the two satellites have present-day stellar mass $(1-2)\times 10^7$, consistent with the observed stellar mass of Fornax dSph, $2\times 10^7\Msun$. Our model predicts the median of 7 and 6 GCs with $M>3\times 10^4\Msun$, with the interquartile ranges spanning $\Ngc = 5-9$ and $4-8$ for \texttt{R2} and \texttt{J2}, respectively. These values match the observations of 6 GCs in Fornax dSph \citep{pace_spectroscopic_2021}.

The masses (both halo and stellar) of the two satellite galaxies grow rapidly over the first 3 Gyr. During this period, the \texttt{R2} galaxy has a smoother mass growth history compared to \texttt{J2}, which has more discrete jumps indicating more frequent major mergers. To show this, we plot in Fig.~\ref{fig:evolv_plots} the mass growth histories of their own satellites\footnote{Since \texttt{R2} and \texttt{J2} are satellites of the main galaxies, these satellite galaxies are satellites of satellites.} that contribute at least 1 surviving cluster to the present-day GC population. The \texttt{R2} galaxy has encountered 2 major mergers both with peak mass $\Mh < 10^{10}\Msun$, whereas \texttt{J2} galaxy has 4 major mergers, and 1 of them has peak mass greater than $10^{10}\Msun$. Around $5\%$ and $20\%$ of surviving GCs are accreted onto \texttt{R2} and \texttt{J2} via mergers, respectively. Although this ratio is small compared to that of MW-size galaxies \citep{chen_modeling_2022}, different merger histories can significantly alter the radial distribution of GCs as we show later.

The two satellite galaxies stop growing mass at a lookback time around 10 Gyr when they accrete onto the central galaxy. The formation of GCs in the two satellites is also quenched at this epoch. After that, the satellite galaxies lose a significant fraction of their halo mass until the present-day. The number of GCs also drops by a factor around 5 compared to the peak value as a result of tidal disruption. 

\begin{figure}
    \includegraphics[width=\linewidth]{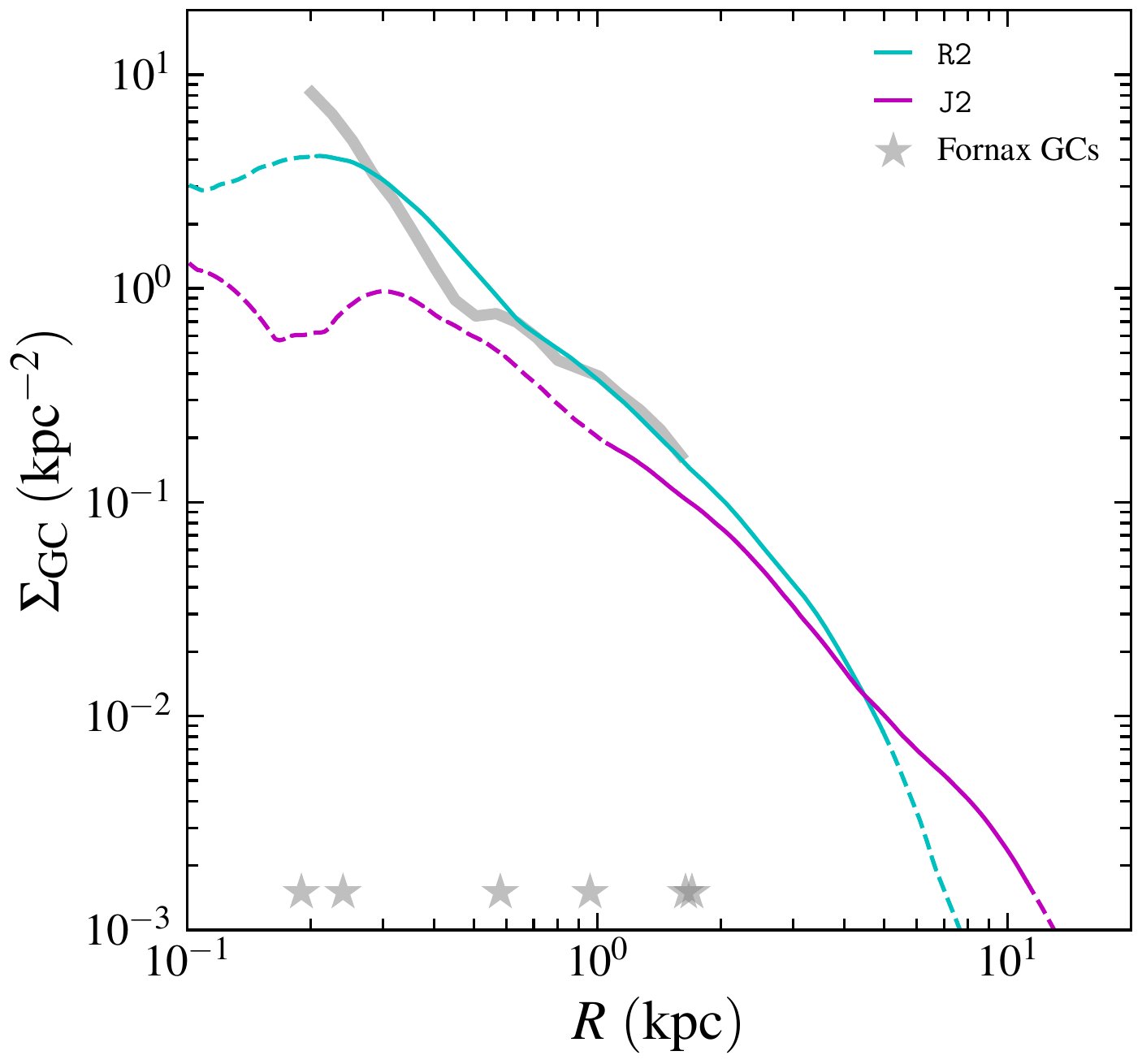}
    \vspace{-4mm}
    \caption{GC number density profiles of the two Fornax-like dwarf galaxies in the \texttt{R\&J} run. The projected radii of the 6 Fornax GCs (Fornax 1--6) to the center of the Fornax dSph are shown as stars. We also show the number density profile of Fornax GCs as the gray curve. The portions with number density below 1 GC per bandwidth are shown as dashed curves.}
    \label{fig:radial_profile}
\end{figure}

We show the GC number density profiles of the two satellites in Fig.~\ref{fig:radial_profile}. The profiles are obtained via kernel density estimation with the \citet{epanechnikov_non-parametric_1969} kernel. The bandwidth of the kernel is 0.3 dex; varying the bandwidth between 0.2 to 0.5 dex does not change the profiles significantly. We also take the observed coordinates of the 6 Fornax GCs from \citet[][Fornax 1-5]{mackey_surface_2003} and \citet[][Fornax 6]{pace_spectroscopic_2021} to compute the projected distances from the center of the Fornax dSph. The \texttt{R2} galaxy can match the observed profile in the radius range where the Fornax GCs are present, $R\lesssim1.6$ kpc. Different from the centrally concentrated GC system of Fornax, \texttt{R2} still hosts GCs out to $R\gtrsim5$ kpc. These GCs raise the half-number radius of \texttt{R2}, 1.3 kpc, to be greater than that of the Fornax dSph, 0.8 kpc. Note that the half-light radius of Fornax is only $\sim 0.8$ kpc \citep{wang_morphology_2019}, and a cluster at $R\gtrsim5$ kpc may not be identified as a member of the galaxy in observations. If we take this selection effect into account and apply a smaller search radius for \texttt{R2}, we can obtain a radial distribution more similar to the Fornax system.

The GC distribution in the more merger-dominated \texttt{J2} galaxy is even more extended, with the half-number radius of 3.2 kpc. \texttt{J2} has lower GC number density than \texttt{R2} for $R\lesssim5$ kpc, but higher in the outside. The \texttt{J2} galaxy has GC number density lower than the statistical significance level (one GC per bandwidth) within the central 1 kpc, whereas it can host GCs out to $R\gtrsim10$ kpc. The GC system in \texttt{J2} is extended likely because major mergers can add kinetic energy to GCs and bring them outwards. It is notable that although \texttt{R2} and \texttt{J2} have similar properties in many aspects, such as the halo mass, stellar mass, and distance to the central galaxies, the GC number density profiles of the two galaxies still differ significantly. If we apply a smaller search radius, the distinct GC distributions in the two galaxies can lead to a notable difference in $\Ngc$. For example, a smaller search radius of 5 kpc does not change $\Ngc$ for \texttt{R2} (recall that the default search radius is the tidal radius $\sim 20$ kpc), but reduces that for \texttt{J2} to $\Ngc=2-5$. 

As mentioned before, both observations and the model show large scatter in $\Ngc$ when scaled with $\Mstar$ or $\Mh$. Here, we suggest that `hidden variables' like the merger history can also alter the observed number of GCs. Although different merger histories do not directly change the number of GCs as mergers only contribute $5-20\%$ of surviving GCs to the two model galaxies, in agreement with the findings that the formation of dwarf galaxies is not dominated by hierarchical assembly \citep{fitts_no_2018,martin_role_2020}, a more merger-rich assembly history may lead to a more extended GC spatial distribution and hence a smaller $\Ngc$ within a fixed search radius.

\section{Constraining model variants}
\label{sec:comparing_settings}

In this section, we compare three alternate model variants to investigate their influences on the $f_{\rm occ}$--$\Mstar$ and $\Ngc$--$\Mstar$ relations. The first alternate model setting employs different lower mass limits $M_{\rm low}$ when counting GCs. By default, we set $M_{\rm low}=3\times 10^4\Msun$ to mimic the $M_{\rm g}<-5.5$ magnitude cut employed in the observations of LV dwarf galaxies \citep{carlsten_elves_2022}. Here, we introduce a lower mass limit of $M_{\rm low}=10^4\Msun$ and a higher mass limit of $M_{\rm low}=10^5\Msun$ to study selection effects due to the cut in GC mass.

Next, we employ an alternate method when sampling cluster mass from the ICMF. As mentioned in Sec.~\ref{sec:cluster_sampling}, by default we sample GC mass from $M_{\rm min}=10^4\Msun$ to $M_{\rm max}\rightarrow\infty$, i.e., there is no higher mass constraint when forming GCs. We thus refer to this setting as `without $M_{\rm max}$' in the subsequent text. In contrast, the previous versions of the model \citep{choksi_formation_2019, chen_modeling_2022} set $M_{\rm max}$ to be a finite value, which is selected to match the deterministic constraint in equation~(\ref{eq:mmax_constraint}). In this setting, a galaxy with mass $M_{\rm h}\sim 10^9\Msun$ has $M_{\rm max}\sim 10^5\Msun$. Therefore, the formation of high mass clusters with $M > 10^5\Msun$ is strictly prohibited in low-mass galaxies. We refer to such a setting as `with $M_{\rm max}$' in the following description.

The third variant explores dependence of the tidal disruption rate on cluster mass, which still remains uncertain. This motivates us to examine the performance of different prescriptions of tidal disruption during GC evolution. The default prescription, as mentioned in Sec.~\ref{sec:cluster_evolution} and equation~(\ref{eq:disrupt}), sets $x=2/3$ and $y=4/3$. Additionally, this prescription approximates the angular frequency $\Omega_{\rm tid}$ in equation~(\ref{eq:t_tid}) by $\Omega_{\rm tid}^2\simeq\lambda_{\rm 1,e}\simeq\lambda_1-\lambda_3$, where $\lambda_{\rm 1,e}$ is the effective tidal strength. We also employ a boost parameter $\kappa$ to account for numerical bias when estimating the tidal tensor: we multiply the derived $\Omega_{\rm tid}^2$ by $\kappa$. By comparing the three MW-like galaxies in the simulations with the observed MW GC system, in Sec.~\ref{sec:selecting_model_parameters} we calibrate $\kappa$ as well as two other model parameters to be $(p_2,p_3,\kappa) = (14,0.7\ {\rm Gyr^{-1}}, 1.5)$. Here, we test two additional prescriptions with $x=y=2/3$ and $x=y=1$. The first of them was applied in our previous work \citep{chen_modeling_2022}. In order to properly compare the three prescriptions, we re-calibrate the model parameters for the alternative prescriptions as in Sec.~\ref{sec:selecting_model_parameters}. After re-calibration, we find $(p_2,p_3,\kappa) = (14,0.7\ {\rm Gyr^{-1}}, 2.5)$ for the $x=y=2/3$ prescription and $(p_2,p_3,\kappa) = (14,0.7\ {\rm Gyr^{-1}}, 1.5)$ for $x=y=1$. The latter parameter set is the same as our fiducial.

\begin{figure}
    \includegraphics[width=\linewidth]{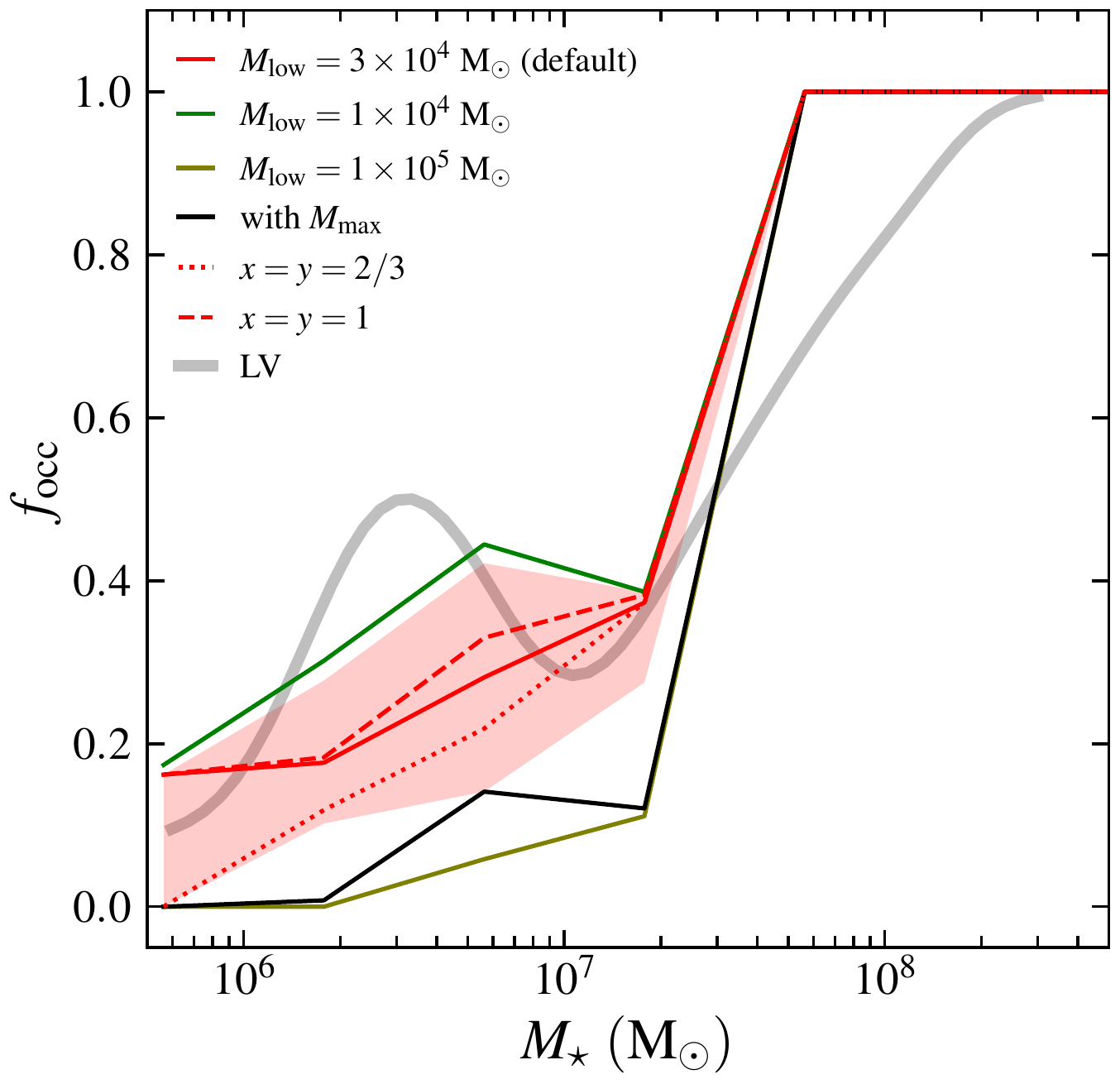}
    \vspace{-4mm}
    \caption{GC occupation fraction $f_{\rm occ}$ as a function of stellar mass of host galaxy $\Mstar$ for different model settings. We show the default model ($M_{\rm low}=3\times10^4\ {\rm M_\odot}$, without $M_{\rm max}$, $x=2/3$, $y=4/3$) as red solid curve with the shaded region representing the interquartile range, in consistency with Fig.~\ref{fig:occupation_frac}. Other models are shown in curves with different styles and colors as described in the legend. The $\Ngc$--$\Mstar$ relation for the LV dwarf galaxies is over-plotted as the gray curve as in Fig.~\ref{fig:occupation_frac}.}
    \label{fig:occupation_frac_compare}
\end{figure}

In Fig.~\ref{fig:occupation_frac_compare}, we compare the occupation fraction predicted by the different model variants. We use the same kernel smoothing method as in Sec.~\ref{sec:number_of_globular_clusters} to make the $f_{\rm occ}$--$\Mstar$ curves. For clarity we show only the $f_{\rm occ}$--$\Mstar$ relations using the $p_{\rm lin}$ distribution function, since the two distribution functions $p_{\rm flat}$ and $p_{\rm lin}$ give consistent results.

We find that $f_{\rm occ}$ varies significantly with the lower mass cut $M_{\rm low}$. Greater $M_{\rm low}$ can significantly reduce $f_{\rm occ}$ in a broad mass range of satellite galaxies from $\Mstar=5\times 10^5$ to $5\times10^7\Msun$. The occupation fractions from the $M_{\rm low}=10^4\Msun$ and $10^5\Msun$ cases differ by around $0.3$ at $\Mstar=10^7\Msun$. We would expect $f_{\rm occ}$ to be invariant of $M_{\rm low}$ if the galaxies host at least one cluster that is more massive than any of the $M_{\rm low}$ employed here. In contrast, such a strong variation suggests that a large fraction of dwarf galaxies can only host GCs less massive than $\lesssim 10^5\Msun$. In fact, among the $20\times25=500$ model satellites (satellites from different realizations are treated as independent galaxies), only $25\%$ can host GCs more massive than $10^5\Msun$, $38\%$ can host GCs more massive than $3\times10^4\Msun$, and $50\%$ can host GCs more massive than $10^4\Msun$.

We also find that the occupation fraction in the model with $M_{\rm max}$ is significantly lower than that in the model without $M_{\rm max}$, for galaxies with $\Mstar\lesssim 5\times 10^7\Msun$. This is because the model with $M_{\rm max}$ strictly prevents the formation of massive clusters with $M\gtrsim 10^5\Msun$ in dwarf galaxies with $M_{\rm h}\lesssim 10^9\Msun$. Clusters initially less massive than $10^5\Msun$ are unlikely to survive tidal disruption to the present-day. However, the model without $M_{\rm max}$ has a small but non-zero probability of forming such massive clusters. In our cluster formation scenario, a galaxy may experience multiple cluster formation events. The cumulative probability of forming at least one massive cluster becomes significant for the model without $M_{\rm max}$ and thus leads to the noticeable difference between the two models. This effect is less important for galaxies with $\Mstar>5\times 10^7\Msun$ as $M_{\rm max}$ becomes large enough to enable the formation of massive clusters. The model with $M_{\rm max}$ predicts the occupation fraction very similar to that in the model with a higher minimum cluster mass $M_{\rm low}=10^5\Msun$.

Finally, the alternate prescriptions of tidal disruption do not change the $f_{\rm occ}$--$\Mstar$ relation noticeably. The occupation fractions from the two alternate prescriptions are always within the interquartile range of the default model. 

To quantitatively evaluate which model agrees better with the observations, we compute the \rms difference between the model and LV satellite galaxies:
\begin{equation}
    \sqrt{\frac{1}{N_j}\sum_j\left(f_{{\rm occ},j}^{\rm model} - f_{{\rm occ},j}^{\rm obs}\right)^2}
    \label{eq:rms}
\end{equation}
where index $j$ stands for the $j$-th mass bin, and $N_j$ is the number of bins. The mass bins are equally spaced in the $\log \Mstar$ space from $\Mstar=10^{5.5}$ to $10^{8.5}\Msun$. We take the bin width to be $0.5$ dex, in consistency with the $f_{\rm occ}$--$\Mstar$ curves for model galaxies in Fig.~\ref{fig:occupation_frac}. The occupation fraction within a bin is calculated with the kernel smoothing method given by equation~(\ref{eq:weighted_value}). For the default model setting ($M_{\rm low}=3\times10^4\ {\rm M_\odot}$, without $M_{\rm max}$, $x=2/3$, $y=4/3$), the \rms deviation is 0.164. For the alternate settings, we find the \rms deviation to be 0.139 for $M_{\rm low}=10^4\Msun$, 0.268 for $M_{\rm low}=10^5\Msun$, 0.249 for the sampling method with finite $M_{\rm max}$, 0.187 for the disruption method of $x=y=2/3$, and 0.157 for $x=y=1$, suggesting that the lower mass limit model with $M_{\rm low}=10^4\Msun$ and the disruption model with $x=y=1$ can match the observed $f_{\rm occ}$--$\Mstar$ relation slightly better than the default model. However, as we show later, the two models perform worse than the default model in matching the $\Ngc$--$\Mstar$ relation. 

\begin{figure}
    \includegraphics[width=\linewidth]{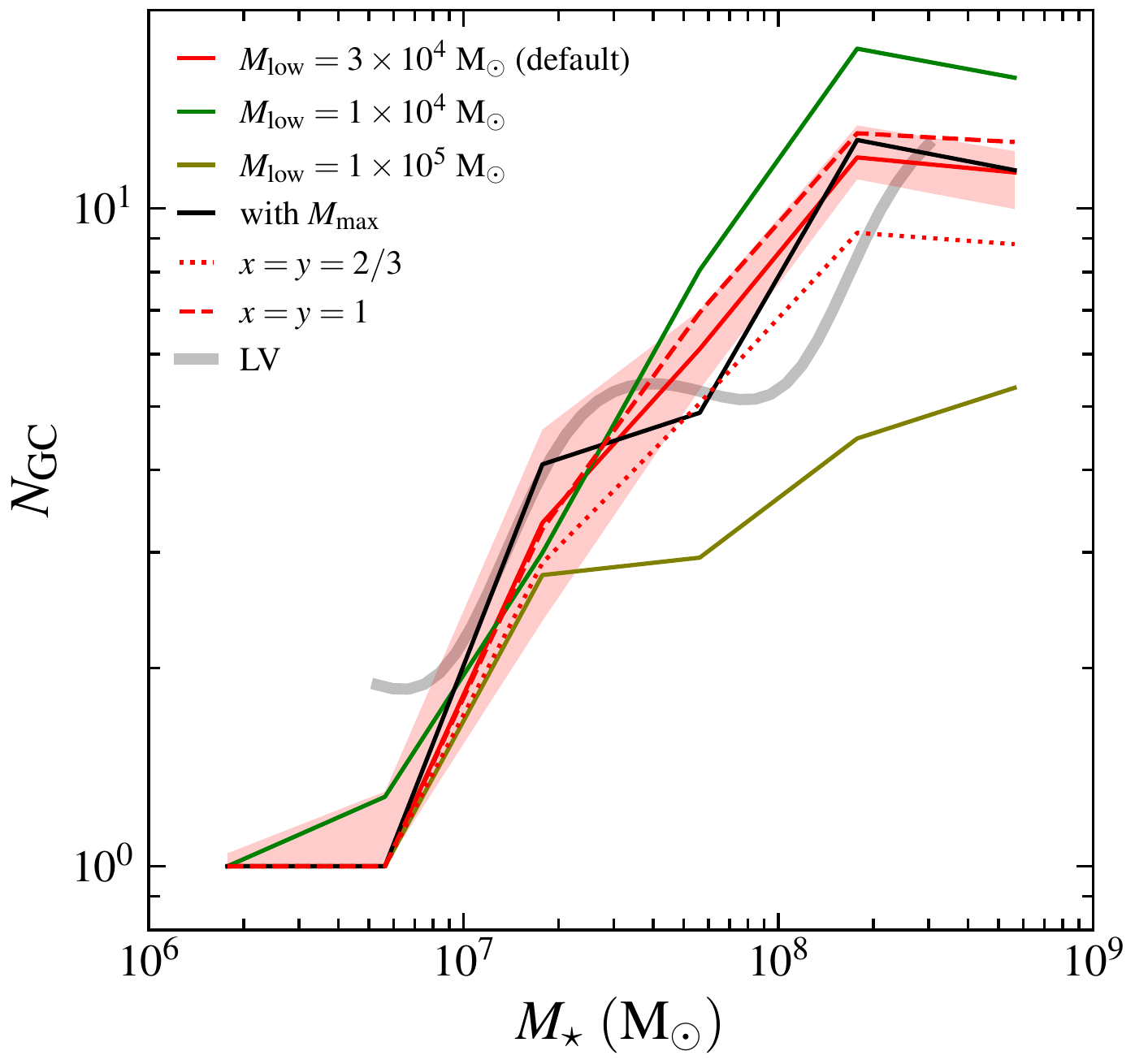}
    \vspace{-4mm}
    \caption{Number of GCs $\Ngc$ as a function of stellar mass of host galaxy $\Mstar$ for different model settings. We show the default model ($M_{\rm low}=3\times10^4\ {\rm M_\odot}$, without $M_{\rm max}$, $x=2/3$, $y=4/3$) as red solid curve with the shaded region representing the interquartile range, in consistency with Fig.~\ref{fig:Ngcs_Ms}. Other models are shown in curves with different styles and colors as described in the legend. The $\Ngc$--$\Mstar$ relation for the LV dwarf galaxies is over-plotted as the gray curve as in Fig.~\ref{fig:Ngcs_Ms}.}
    \label{fig:Ngcs_Ms_compare}
\end{figure}

Next, we show in Fig.~\ref{fig:Ngcs_Ms_compare} the number of GCs as a function of $\Mstar$ for different model variants. Before going deep into the analysis, we emphasize that it is meaningless to look at the $\Ngc$--$\Mstar$ relation below $\Mstar\simeq5\times10^6\Msun$ because these low-mass galaxies normally host at most 1 GC, and usually none. Since we analyse here galaxies with non-zero GCs, the $\Ngc$ value is almost always 1 regardless of model settings. It is therefore more meaningful to look at the occupation fraction mentioned before for galaxies below $\Mstar\simeq5\times10^6\Msun$.

We note that $\Ngc$ is sensitive to $M_{\rm low}$. As $M_{\rm low}$ increases from $10^4$ to $10^5\Msun$, the number of GCs in a galaxy can drop by a factor of 3 at $\Mstar\simeq 10^8\Msun$. In contrast, $\Ngc$ is not greatly affected by the alternate sampling model with $M_{\rm max}$, except for a more wiggled $\Ngc$--$\Mstar$ relation, although this model predicts significantly lower $f_{\rm occ}$ at $\Mstar=5\times10^5-5\times10^7\Msun$. This is because the `without $M_{\rm max}$' and `with $M_{\rm max}$' models become equivalent when the galaxy is massive enough to host at least one massive cluster that can survive to the present-day. Finally, the $\Ngc$--$\Mstar$ relation is almost unchanged with the alternate disruption prescriptions.

Again, we quantitatively evaluate model agreement with observations by computing the \rms deviation between the model and LV $\Ngc$:
\begin{equation}
    \sqrt{\frac{1}{N_j}\sum_j\left(\frac{N_{{\rm GC},j}^{\rm model} - N_{{\rm GC},j}^{\rm obs}}{\sigma_{N,j}}\right)^2}.
    \label{eq:rms_n}
\end{equation}
Here we include the denominator $\sigma_N$ to account for the uncertainties in the number of GCs. We set $\sigma_{N,j}=\sqrt{N_{{\rm GC},j}^{\rm obs}}$ as the Poisson's error when counting GCs. The mass bins are equally spaced in the $\log \Mstar$ space from $\Mstar=10^{6.5}$ to $10^{8.5}\Msun$. For the default model setting the \rms deviation is 0.725. For the alternate settings, we find the \rms deviation to be 1.702 for $M_{\rm low}=10^4\Msun$, 0.955 for $M_{\rm low}=10^5\Msun$, 0.806 for the sampling method with finite $M_{\rm max}$, 0.434 for the disruption method of $x=y=2/3$, and 0.940 for $x=y=1$. Only the $x=y=2/3$ model performs better than our default model in matching the occupation fraction. Although the $M_{\rm low}=10^4\ {\rm M_\odot}$ case can match the occupation fraction slightly better, it significantly overestimates the number of GCs. The alternate disruption prescriptions have similar performance to the default model: the $x=y=2/3$ model gives a better match for $\Ngc$ but a worse match for $f_{\rm occ}$, while the $x=y=1$ model gives a better match for $f_{\rm occ}$ but a worse match for $\Ngc$. Since the default setting $x=2/3,y=4/3$ can better match the observed GC mass function of the MW (see Fig.~\ref{fig:mass_function}), we favor the default setting over the alternate disruption models.

\section{Summary and Discussion}
\label{sec:summary}

In this work, we tested the performance of the GC formation and evolution model in the dwarf galaxy regime ($\Mh<10^{11}\Msun$) resembling the LG environment. The model is based on our previous work \citep{chen_modeling_2022} with four stages: cluster formation, cluster sampling, particle assignment, and cluster evolution. We use empirical scaling relations to calculate the total mass of GCs from the halo merger history, and stochastically sample the mass of individual clusters. We have removed the deterministic setting of maximum GC mass in the previous sampling method to allow the formation of massive clusters ($M\gtrsim10^5\Msun$) in low-mass galaxies ($\Mh\lesssim10^9\Msun$) with a small but non-zero probability. Such stochasticity is important for correctly reproducing the observed GC occupation fraction in dwarf galaxies. Different from our previous work, where we preferentially assign GCs to stellar particles, here we assign GCs to DM particles restricted to local density peaks because we use collisionless simulations of the LG environment with sufficiently high mass resolution to capture all relevant dwarfs. We additionally require all GCs to form within the scale radius of the host galaxy. These settings ensure resulting radial number density profiles of model GCs can match the observed profiles of both the MW and satellites. We also employ a new prescription of tidal disruption that produces stronger disruption of low-mass clusters and can better match the observed GC mass function. 

Despite these minor adjustments, the overall structure of the model remain unchanged: the model still has three adjustable parameters that control the formation rate, formation timing, and disruption rate of GCs. It is worth noting that although the main focus of this work is the LG dwarf satellites, we only calibrate the model parameters by comparing key properties of the GC systems in the three central MW-like galaxies from the simulations to the observed properties of the MW GC system. Therefore, any consistency with observations of dwarf GC systems is a true prediction of the model rather than an outcome of fitting the data.

We run the calibrated model on 20 satellite galaxies in the simulated LG systems and repeat 25 times with different random seeds to study how much the resulting GC systems are influenced by the model randomness. Since the central galaxy may tidally strip GCs from the host satellite if the GCs are too distant from the satellite, we only count GCs within the tidal radius of the satellite galaxy. Our model performs surprisingly well in matching the occupation fraction and number of GCs in the dwarf regime with the LV observations by \citet{carlsten_elves_2022}, see Figs.~\ref{fig:occupation_frac}, \ref{fig:Ngcs_Ms}, and \ref{fig:Ngcs_Ms_wide}. This consistency implies that the physics of GC formation and evolution may be universal for both central and satellite galaxies. 

Dwarf galaxies in this study can only host a few or even no GCs. Small number statistics becomes important as a minor change in any physical process that is relevant to GC formation or evolution may introduce significant variance in the number of GCs. In an even lower-mass regime $\Mstar\lesssim10^7\Msun$, most galaxies host less than 2 GCs, with cluster mass $\lesssim 10^5\Msun$. The ability to match the observed number of GCs and the occupation fraction in such a regime is a very strict test of the model implementation of cluster formation and disruption mechanisms.

We also test different model settings to study their influence on the observable results. We find that the occupation fraction statistic primarily constrains the low-mass cut when counting GCs, $M_{\rm low}$, and the potential existence of maximum GC mass, $M_{\rm max}$ (Fig.~\ref{fig:occupation_frac_compare}). Since most dwarf galaxies in this study can only host clusters with mass $\lesssim 10^5\Msun$, $M_{\rm low}$ varying from $10^4$ to $10^5\Msun$ can cause the occupation fraction to differ by $\sim0.3$. In addition, the `with $M_{\rm max}$' model strictly prevents the formation of massive clusters ($M\gtrsim10^5\Msun$) in low-mass galaxies ($\Mh\lesssim10^9\Msun$) by setting a deterministic upper mass limit. Although this setting is not very different from the default `without $M_{\rm max}$' model in massive galaxies ($\Mh>10^9\Msun$), it predicts a much lower occupation fraction in the low-mass end where majority of clusters are tidally disrupted.

On the other hand, the GC number statistic sets strong constraints on $M_{\rm low}$ but is not sensitive to $M_{\rm max}$ (Fig.~\ref{fig:Ngcs_Ms_compare}). As $M_{\rm low}$ increases from $10^4$ to $10^5\Msun$, $\Ngc$ drops by a factor of $\sim 3$ at $\Mstar\simeq10^8\Msun$. Since the average GC mass in low-mass galaxies is typically lower than in the MW-size galaxies, it is important to correct for the observational incompleteness below the detection limit in dwarf GC systems. In contrast, the `with $M_{\rm max}$' model predicts a similar $\Ngc$--$\Mstar$ relation compared to the default `without $M_{\rm max}$' model since the two models become equivalent when the galaxy is massive enough to host massive clusters that can survive to the present-day. 

We also investigate the near-linear $\Mgc$--$\Mh$ relation in a wide mass range across 6 orders of magnitude (Fig.~\ref{fig:Ngcs_Mh}). By jointly fitting the observational data from \citet{harris_galactic_2017} and \citet{forbes_extending_2018} we obtain $\Ngc\propto\Mh^{0.77\pm0.03}$ and $\Mgc\propto\Mh^{0.93\pm0.03}$ with a significant intrinsic scatter of $0.3-0.4$~dex. The latter relation is very close to linearity, and has been reliably confirmed for $\Mh=10^{10}-10^{15}\Msun$. Since independently measuring the halo mass is challenging below $\Mh\simeq10^{10}\Msun$, only a limited number of works attempted to study the relation in the low-mass end. Our model predicts an $\Mgc$--$\Mh$ relation in agreement with the observational relation in the low-mass end down to $\Mh\simeq10^8\Msun$. We emphasize that since our model is only calibrated for the central galaxies, such an agreement indicates that the near-linear $\Mgc$--$\Mh$ relation is an evidence for universal physical processes governing GC formation and evolution in galaxies of all size, from dwarfs to giants.

We discuss a specific example of two satellite GC systems similar to that of Fornax dSph, which previously remained unexplained. We find the systems could have contained up to $30-50$ GCs in the past but have stopped GC formation after accretion onto the central galaxies $\sim10$ Gyr ago (Fig.~\ref{fig:evolv_plots}). There are two mechanisms that reduce the number of satellite GCs: tidal disruption and tidal stripping by the central galaxy. We find that our two example galaxies have already had peri-galactic encounters prior to the present. Only $4-9$ GCs in the two galaxies can survive and remain inside the tidal radius. 

We note that GCs in the two Fornax-like galaxies are located out to $5-10$ kpc (Fig.~\ref{fig:radial_profile}), which is much larger than the effective radius of Fornax dSph, $0.8$ kpc. Observationally, these GCs are unlikely to be identified as members of the galaxy since they are too distant. To avoid the biased measurement of $\Ngc$ observations must employ a large enough search radius ($\gtrsim r_{\rm tid}$). However, this is challenging as the background GCs may be indistinguishable from the GCs belonging to the satellite at such a large radius.

Moreover, different merger history can also alter the radial distribution of GCs: the more merger-dominated satellite has an even more extended GC system. The satellite with fewer major mergers better matches the GC number density profile in the Fornax dSph. Although different merger histories may not directly change the number of GCs, a more merger-rich assembly history leads to a more extended GC spatial distribution and hence a smaller $\Ngc$ within a fixed search radius. This is one of the `hidden variables' that contributes to the scatter in the $\Ngc$--$\Mstar$ and $\Ngc$--$\Mh$ relations.

\section*{Acknowledgements}
We thank Gillen Brown, Eric Bell, Mark Gieles, Xi Meng, Leandro Beraldo e Silva, and Monica Valluri for insightful discussions. This research was mainly conducted with the \textsc{python} programming language, employing the following packages: \textsc{numpy} \citep{harris_array_2020}, \textsc{matplotlib} \citep{hunter_matplotlib_2007}, \textsc{yt} \citep{turk_yt_2011}, \textsc{scipy} \citep{virtanen_scipy_2020}, and \textsc{astropy} \citep{the_astropy_collaboration_astropy_2018}. OG and YC were supported in part by the U.S. National Science Foundation through grant AST-1909063 and by National Aeronautics and Space Administration through contract NAS5-26555 for Space Telescope Science Institute program HST-AR-16614.

\section*{Data Availability}

The data that support the findings of this study are available from the corresponding author upon reasonable request. 


\bibliographystyle{mnras}
\bibliography{GC-model-references}

\bsp	
\label{lastpage}
\end{document}